\documentclass[
superscriptaddress,
twocolumn,
nofootinbib,
 amsmath,amssymb,
 aps,
prd,
]{revtex4-2}

\pdfoutput=1
\usepackage{natbib}
\usepackage{aas_macros}
\usepackage{lipsum}
\usepackage{ulem}
\usepackage{graphicx}
\usepackage{dcolumn}
\usepackage{bm}
\usepackage{hyperref}
\usepackage{booktabs}
\usepackage{color}
\usepackage{hyperref}
\hypersetup{
    colorlinks=true,
    linkcolor=blue,
    filecolor=blue,      
    urlcolor=blue,
    citecolor=blue,
    }

\newcommand{\be}{\begin{equation}}
\newcommand{\ee}{\end{equation}}
\newcommand{\barr}{\begin{eqnarray}}
\newcommand{\earr}{\end{eqnarray}}
\newcommand{\bal} {\begin{aligned}}
\newcommand{\eal} {\end{aligned}}
\newcommand{\bl}{\boldsymbol{l}}
\newcommand{\bL}{\boldsymbol{L}}

\newcommand{\bll}{\boldsymbol{L}}

\newcommand{\dotfac}[1]{({\bll} \cdot {\bl}_{#1})}

\newcommand{\intL}{\int_{\substack{\bl_1 + \bl_2 \\ =\bll }}}

\begin{document}

\preprint{APS/123-QED}

\title{Constraining Cosmology With the CMB \texorpdfstring{$\times$}{TEXT} LIM-Nulling Convergence}

\author{Hannah Fronenberg}
\email{hannah.fronenberg@mail.mcgill.ca}
\affiliation{Department of Physics, McGill University, 3600 Rue University, Montreal, QC H3A 2T8, Canada}
\affiliation{Trottier Space Institute, 3550 Rue University, Montreal, QC H3A 2A7, Canada}
\author{Abhishek S. Maniyar}
\affiliation{Department of Physics, Stanford University, 382 Via Pueblo Mall, Stanford, CA 94305, USA}
\affiliation{SLAC National Accelerator Laboratory, 2575 Sand Hill Rd, Menlo Park, CA 94025, USA}
\affiliation{Kavli Institute for Particle Astrophysics and Cosmology, 382 Via Pueblo Mall Stanford, CA  94305-4060, USA}%
\author{Anthony R. Pullen}
\affiliation{Department of Physics, New York University, 726 Broadway, New York, NY, 10003, USA}%
\affiliation{Center for Computational Astrophysics, Flatiron Institute, New York, NY 10010, USA}%
\author{Adrian Liu}
\affiliation{Department of Physics, McGill University, 3600 Rue University, Montreal, QC H3A 2T8, Canada}
\affiliation{Trottier Space Institute, 3550 Rue University, Montreal, QC H3A 2A7, Canada}

\date{\today}
\begin{abstract}

Lensing reconstruction maps from the cosmic microwave background (CMB) provide direct observations of the matter distribution of the universe without the use of a biased tracer. Such maps, however, constitute projected observables along the line of sight that are dominated by their low-redshift contributions. To cleanly access high-redshift information, Ref.~\cite{Maniyar_nulling} showed that a linear combination of lensing maps from both CMB and line intensity mapping (LIM) observations can exactly null the low-redshift contribution to CMB lensing convergence. In this paper we explore the scientific returns of this nulling technique. We show that LIM-nulling estimators can place constraints on standard $\Lambda$CDM plus neutrino mass parameters that are competitive with traditional CMB lensing. Additionally, we demonstrate that as a clean probe of the high-redshift universe, LIM-nulling can be used for model-independent tests of cosmology beyond $\Lambda$CDM and as a probe of the high-redshift matter power spectrum.

\end{abstract}

\maketitle


\section{\label{sec:intro}Introduction}

In recent decades, there has been a sustained effort to make precision measurements of the large scale universe over a vast portion of its history. Line intensity mapping (LIM) is an emergent technique for studying large scale structure. Here, one observes the integrated intensity of a single spectral line emanating from galaxies and the intergalactic medium (IGM). By virtue of observing lines with known rest frequencies, line intensity mapping allows one to obtain precise redshift information. Mapping line emission over a large bandwidth can therefore yield unprecedentedly large maps of the universe in three dimensions, allowing us to observe cosmic evolution in action. A number of lines are being targeted by current and upcoming experiments including Lyman-$\alpha$, H-$\alpha$, the 21 cm line of neutral hydrogen (HI), the 3727 {\AA}, 3729 {\AA} lines of singly ionized oxygen ([OII]), the forbidden 88.4 $\mu$m and 51.8 $\mu$m transitions of doubly ionized oxygen ([OIII]), a host of rotational line transitions of carbon monoxide (CO), and the forbidden 158 $\mu$m line of ionized carbon ([CII]). Each line traces a biased matter density field as well as regions of the IGM and of the galaxy related to their specific emission or absorption mechanisms. This makes line intensity mapping a powerful probe of both cosmology and astrophysics. 

Along with LIMs, gravitational lensing is a promising probe of the matter density field. Weak gravitation lensing of the cosmic microwave background (CMB) arises when CMB photons from the surface of last scattering get deflected by the gravitational potentials that they encounter on their journey to the observer. Using CMB temperature and polarization maps to reconstruct the lensing potential, $\phi$, gives us direct observation of the total matter distribution of the universe, both baryonic and dark, without the use of a biased tracer \cite{Zaldarriaga_1998}. Measuring the power spectrum of the lensing potential, either in auto- or in cross-correlation with large scale structure surveys, has the ability to probe the growth of matter fluctuations, place limits on primordial non-Gaussianity, constrain the sum of the neutrino masses, and even test theories of modified gravity \cite{Lewis_Challinor2007,Schmittfull_2018,Allison_2015}. This information, however, is collapsed onto a single observable, the convergence,  and the high-redshift contribution to the convergence is dwarfed by that of the low-redshift universe ($z \lesssim 2$). Since the CMB lensing convergence contains information about how matter is distributed along the entire line of sight, it has the potential to help us trace out the matter distribution of the early universe. 

There are several proposed techniques to disentangle the redshift integrated lensing signal and to extract information from particular redshift intervals. For instance, cross-correlating the CMB convergence field with another tracer, such as a galaxy survey or line intensity map, allows one to pick out common matter density correlations a their common redshift. This method, however, has its drawbacks. The redshifts available for study are limited to those of the non-lensing probe and by virtue of cross-correlating with a biased tracer, the resulting cross-correlation is likewise biased, losing out on the unbiased nature of the lensing convergence. One can make progress on the latter by considering not a correlation with the tracer itself but rather, for example, with the LIM lensing convergence. Just like the CMB, LIMs also experience weak lensing by large scale structure as the photons pass through the cosmos on their way to our instruments. These lines, however, are only lensed by a portion of large scale structure that lenses the CMB, namely the low redshift universe. Cross-correlating LIM lensing and CMB lensing allows one to study the common low redshift matter density field that lenses both the LIM and the CMB. While the resulting correlation is unbiased it again is limited to the redshifts between the observer and the source plane of the LIM.

In order to access the high redshift information,  Ref.~\cite{Maniyar_nulling} proposes using the lensing information of two LIMs, to not just suppress, but exactly null out the low redshift contribution to the CMB convergence. This ``nulling" method has been explored in the context of galaxy lensing \cite{Huterer_nulling_2005,Bernardeau_nulling_2014,Barthelemy_nulling_2020} as well as CMB lensing \cite{McCarthy_2021,Qu_2023,Zhang_2023}. For instance, Ref. \cite{McCarthy_2021} show that one can ``null" out the imprint of uncertain baryonic effects from CMB lensing maps using cosmic shear surveys at $z < 1$. Similarly, Ref. \cite{Qu_2023} showed that one can also use cosmic shear surveys to subtract off the imprint of uncertain dark energy physics from CMB lensing maps. Ref. \cite{Zhang_2023} explores the potential of subtracting the imprint of gravitational nonlinearity at low redshift to help measure primordial bispectra. While never implemented with real data, these nulling techniques could be an important new tool for studying the high redshift universe. 

What is more, Ref. \cite{Maniyar_nulling} show that the CMB $\times$ LIM-nulling convergence spectrum, $\langle \hat{\kappa}\hat{\kappa}_{\rm{null
}} \rangle$, does not contain so-called line interloper bias when the LIM convergence maps are estimated with ``LIM-pair" estimators of Section~\ref{subsec:LIM_lensing}. Line interlopers are one of the chief systematic contaminants in LIMs, and consist of low-redshift spectral lines that redshift into the same observed frequency channel as the high redshift target line. In addition to estimators such as the LIM-pair estimators that mitigate interloper bias by construction, other strategies such as line identification, analysis of redshift space distortions, spectral deconfusion, and cross-correlations have all been shown to help reduce line interloper contamination \citep{Masking_silva,Sun_2018,Visbal_2011,Masking_Breysse,Gong_2014,2016ApJ...833..242L}. Provided that some combination of these strategies is able to bring line interlopers (and other potential systematics) under control, the CMB $\times$ LIM-nulling convergence spectrum has the potential to reveal exclusive, information about the early universe. Exactly what information is revealed is the subject of this paper. 

In this work, we explore the parameter space of LIM-nulling measurements. In section \ref{sec:LIM_lensing_nulling} we derive, for the first time, the CMB $\times$ LIM-nulling variance as well as discuss the potential use for this probe in constraining cosmology. In a companion paper, Ref. \cite{LIM_nulling_BAO}, we forecast using the CMB $\times$ LIM-nulling convergence spectrum to detect baryon acoustic oscillations in the early universe, which may serve as a standard ruler over vast portion of cosmic history. In Sections \ref{sec:Choice_of_Lines} and \ref{sec:Sensitivity}, calculate the signal-to-noise ratio of this cross-spectrum statistic as a function of various observing parameters. After exploring this vast parameter space, we converge on three possible observing scenarios some of which would allow one to constrain cosmology at high redshift in Section \ref{sec:Sensitivity}. Using these scenarios, we present several forecasts. Section \ref{sec:LCDM Forecast} consists of a series of forecasts on $\Lambda$CDM+$M_{\nu}$ cosmology. First is a Fisher forecast to test the sensitivity of the CMB $\times$ LIM-nulling to the concordance model of cosmology and to compare it with traditional CMB lensing forecasts. In this section, we also explore how this LIM-nulling estimator behaves, in comparison to the regular CMB lensing convergence, in universes with time-evolving cosmologies. In addition, we forecast the sensitivity of this probe to the matter power spectrum at high-$z$ in Section \ref{sec:PCA}. Unless otherwise explicitly stated, our fiducial cosmology is that of \textit{Planck 2015}. 


\section{\label{sec:LIM_lensing_nulling}LIM Lensing and LIM Nulling}

In this section, we outline the key lensing estimators and observables used throughout the text which follow Ref. \cite{Maniyar_nulling}. In the first subsection, we provide a brief overview of weak lensing by large scale structure in the context of the CMB and extend the discussion to include LIMs. In Section \ref{subsec:LIM_lensing}, we 
quickly review LIM lensing estimator and LIM-pair estimators. Finally, in section \ref{subsec:Nulling}, we build upon the existing LIM-nulling estimator formalism to provide a derivation of the CMB $\times$ LIM-nulling variance. 

\subsection{\label{subsec:weak_lensing} Weak Lensing By Large-Scale Structure}

The CMB acts as a source image which is lensed by the intervening matter density field. The deflection angle, $\alpha$, is proportional to the gradient of the lensing potential, $\phi$, which is the total gravitational potential of the projected mass distribution along the line of sight. This gradient of the potential is related to the convergence, $\kappa = -1/2 \nabla \phi$, which is the line-of-sight-integrated matter density field, given by 
\begin{equation}\label{eq:convergence}
    \kappa(\mathbf{\hat{n}}) = \int_0^{z_{\rm{s}}} W(z',z_{\rm{s}})\delta_{m} (\chi(z')\mathbf{\hat{n}},z')\frac{c \, dz'}{H(z')}
\end{equation}
where $z_{\rm{s}}$ is the redshift of the source, $ W(z,z_s)$ is the lensing kernel, $H(z)$ is the Hubble parameter, c is the speed of light, $\chi$ denotes the comoving distance, and $\delta_{m} (\mathbf{r},z)$ is the matter density field at position $\mathbf{r}$ and redshift $z$. The lensing kernel for a source at a single comoving slice is given by 

\begin{equation}\label{eq:lensing_kernel}
    W(z,z_s) = \frac{3}{2}\left(\frac{H_0}{c}\right)^2\frac{\Omega_{m,0}}{a}\chi(z)\left(1-\frac{\chi(z)}{\chi(z_s)}\right)
\end{equation}
where $H_0$ is the Hubble constant, $\Omega_{m,0}$ is the matter fraction today, $a$ is the scale factor, and $\chi(z_s)$ is the comoving distance to the source.


The deflections induced on CMB photons are small, on the order of arcminutes. However, the structures responsible for the deflection are large, on the order of degrees. Therefore, somewhat counter-intuitively, to study the large scale structure of the universe one actually has to study the small scale anisotropies of the CMB. Lensing induces correlations between the otherwise uncorrelated CMB spherical harmonic coefficients, $a_{l m}$. With the use of quadratic estimators like those derived in Refs. \cite{Hu_Okamoto_2002} and \cite{Maniyar_QE}, an estimate of $\kappa$ can be obtained which we denote by $\hat{\kappa}$.

One can then compute the angular power spectrum of the convergence which is given by 
\begin{multline}\label{eq:CMB_spectrum}
    C_{L}^{\hat{\kappa} \hat{\kappa}} = \int_0^{z_s} \frac{W(z',z_{\rm{s}})^2}{\chi(z')^2}P_{\rm{m}}\left(k = \frac{L+1/2}{\chi(z')},z'\right)\frac{c \, dz'}{H(z')}
\end{multline}
where $P_m$ is the matter power spectrum. In this expression, we assume the Limber approximation.

Obtaining lensing measurements are challenging yet have seen tremendous progress in recent years. To date, a number of lensing detections have been made, the first of which was by the Wilkinson Microwave Anisotropy Probe (WMAP) in 2007 using the Hu and Okamoto estimator on temperature maps and cross-correlating the resultant $\kappa$ map with radio galaxy counts \cite{WMAP_lensing}. Subsequent measurements have been made of the lensing signal in temperature as well as polarization maps by ACT \cite{ACT_lensing_2011,ACT_lensing_2017,ACT_lensing_cross_2023}, the South Pole Telescope (SPT)    \cite{SPT_lensing_2012,SPT_lensing_2013,SPT+Planck_lensing_2017,SPTpol_lensing_2015,SPTpol_lensing_2019,SPT3G_2023}, \textit{Planck} \cite{Planck_lensing_2015,Planck_lensing_2018}, Background Imaging of Cosmic Extragalactic Polarization (BICEP) \cite{BICEP_lensing_2016}, and the POLARization of the Background Radiation experiment (POLARBEAR)\cite{POLARBEAR_lensing_2015,POLARBEAR_lensing_2020}. Excitingly, current and next generation wide-field CMB experiments like SPT-3G, SPT-3G+, AdvACT, the Simons Observatory (SO) and CMB-Stage 4 (CMBS4), will provide high signal-to-noise lensing measurement with unprecedented angular resolution \cite{AdvACT,SPT-3G,SPT-3G+,SO,CMB-S4}.  These upcoming detections will enable further analyses such as the LIM-nulling measurement we propose in Section \ref{subsec:Nulling} and forecast in Section \ref{sec:LCDM Forecast}.

\subsection{\label{subsec:LIM_lensing} LIM Lensing Estimators}

Just like the CMB, lower redshift LIMs also incur correlations between Fourier coefficients as a result of lensing. In the same spirit as CMB lensing reconstruction, the LIM lensing convergence can be estimated with LIM lensing estimators which are extensions of those developed for the CMB. LIMs however, suffer from significant foreground bias, be it from diffuse extended sources, or from line interlopers. This has been shown to cause significant foreground bias to the LIM lenisng convergence \cite{Maniyar_nulling}. Luckily, Ref. \cite{Maniyar_nulling} showed that by using a LIM-pair estimator, one can perform LIM lensing reconstruction free of interloper bias (to first order). This LIM-pair estimator makes use of the fact that two LIMs from the same redshift slice and the same patch of the sky will contain the same correlations due to lensing. However, since each line is observed at a different frequency, they will suffer from different sources of foreground contamination which will be uncorrelated. The LIM-pair lensing estimator, using LIMs $X$ and $Y$, is given by 

\be\label{eq:LIM-pair estimator}
\hat{\kappa}_{XY}(\bll) 
= 
\int \frac{d^2 l_1}{(2 \pi)^2}
\frac{d^2 l_2}{(2 \pi)^2} \
F_{XY}(\bl_1, \bl_2) \ X_{\bl_1} Y_{\bl_{2} }\ , 
\ee
where $\bl$ are the two dimensional Fourier wavenumbers for the LIM in the flat sky approximation, and $\bL = \bl_{1} + \bl_{2}$ are the wavenumbers for the lensing potential. The quantities $X(\bl)$ and $Y(\bl)$ are the observed LIM fields in Fourier space. The function $F_{XY}$ is uniquely determined to ensure that $\hat{\kappa}_{XY}$ is unbiased to first order and to ensure that $\hat{\kappa}_{XY}$ is the minimum variance estimate of $\kappa_{XY}$. This solution to $F_{XY}$ is given by 
\be\label{eq:F_XY}
\bal
&F_{XY}(\bl_1, \bl_2) 
= \lambda_{XY}(L) \times\\
&\quad\quad\frac{C_{l_1}^{YY} C_{l_2}^{XX} f_{XY}(\bl_1, \bl_2) - C_{l_1}^{XY} C_{l_2}^{XY}  f_{XY}(\bl_2, \bl_1)}{C_{l_1}^{XX} C_{l_2}^{YY}C_{l_1}^{YY} C_{l_2}^{XX} - \left(C_{l_1}^{XY} C_{l_2}^{XY}\right)^2}  \eal
\ee
where $C_{l}^{XX}$ and $C_{l}^{YY}$ are the total auto-spectra for LIMs $X$ and $Y$ including noise, while $C_{l}^{XY}$ is their cross-spectrum. The Lagrange multiplier $\lambda_{XY}(L)$ is given by 
\be
\bal \label{eq:lagrange_xy}
\lambda_{XY}(L) &\equiv \Bigg[\intL f_{XY}(\bl_1, \bl_2) \times \\ 
&\frac{C_{l_1}^{YY} C_{l_2}^{XX} f_{XY}(\bl_1,\bl_2) - C_{l_1}^{XY} C_{l_2}^{XY} f_{XY}(\bl_2,\bl_1)}{C_{l_1}^{XX} C_{l_2}^{YY}C_{l_1}^{YY} C_{l_2}^{XX} - \left(C_{l_1}^{XY} C_{l_2}^{XY}\right)^2} \Bigg]^{-1}. 
\eal
\ee
where for brevity, we introduce the notation
\be
\intL ... \equiv \iint \frac{d^2 l_1 d^2 l_2}{(2 \pi)^2} \delta(\bl_1 + \bl_2 - \bll) ... \,.
\ee
The factor $f_{XY}(\bl,\bl')$ is the coupling coefficient

\be\label{eq:f_xy}
f_{XY}(\bl, \bl') = 
-\frac{2}{L^2}
\left[
\widetilde{C}_{l_1}^{XY}  \dotfac{1} 
+  \widetilde{C}_{l_2}^{XY}  \dotfac{2} 
\right]
\ee
where $\widetilde{C}_{l}^{XY}$ is the unlensed cross-power spectrum. The reconstruction noise, $N_{XY}(L)$, of the LIM-pair estimator is given by

\be\label{eq:limpair_variance}
\bal
N_{XY}&(L)
=
\intL F_{XY}(\bl_1, \bl_2) \times \\ 
&\Big( F_{XY}(\bl_1, \bl_2) C_{l_1}^{XX} C_{l_2}^{YY} + F_{XY}(\bl_2, \bl_1) C_{l_1}^{XY} C_{l_2}^{XY} \Big).
\eal
\ee

For the interested reader, a detailed derivation of this estimator can be found in Appendix B of Ref. \cite{Maniyar_nulling}.



\subsection{\label{subsec:Nulling} CMB \texorpdfstring{$\times$}{TEXT} LIM-nulling}

In the last two subsections, we outlined how one could make use of both the CMB and of LIMs to extract information about the intervening matter density field. In the case of the CMB the resulting field is the matter density field over cosmic history since the surface of last scattering, projected onto a single plane. It is important to note that LIMs are lensed by the same low-$z$ gravitational potentials that lens the CMB and therefore these probes share common low-redshift induced correlations. This can be exploited in order to make use of the LIM lensing information to ``clean" the CMB convergence of its low redshift contribution. 

From Eq. \eqref{eq:convergence}, it should be clear that it is possible to construct some kernel that vanishes over the low redshift interval $[0,z_{\rm{null}}]$. Since $W$ is quadratic in $\chi$, a linear combination of three such kernels suffices to find a non-trivial null solution for the coefficients of this polynomial. As shown in Ref. \cite{Maniyar_nulling}, using two convergence maps each estimated from two LIMs at redshifts $z_1$ and $z_2$ ($z_{1} < z_{2}$), and one CMB convergence map sourced at the surface of last scattering, $z_{\rm{CMB}}$, the LIM-nulling kernel is given by 
\begin{equation}
\label{eq:nulling_kernel}
    W_{\rm{null}} = W(z,z_{\rm{CMB}}) + \alpha W(z,z_2) - (1+\alpha)W(z,z_1)
\end{equation}
where 
\be\label{eq:alpha}
\alpha \equiv \frac{1/\chi(z_{\rm{CMB}})- 1/\chi(z_1)}{1/\chi(z_1) - 1/\chi(z_2)}.
\ee

\begin{figure}[ht]
\includegraphics[width = 9.5cm]{ 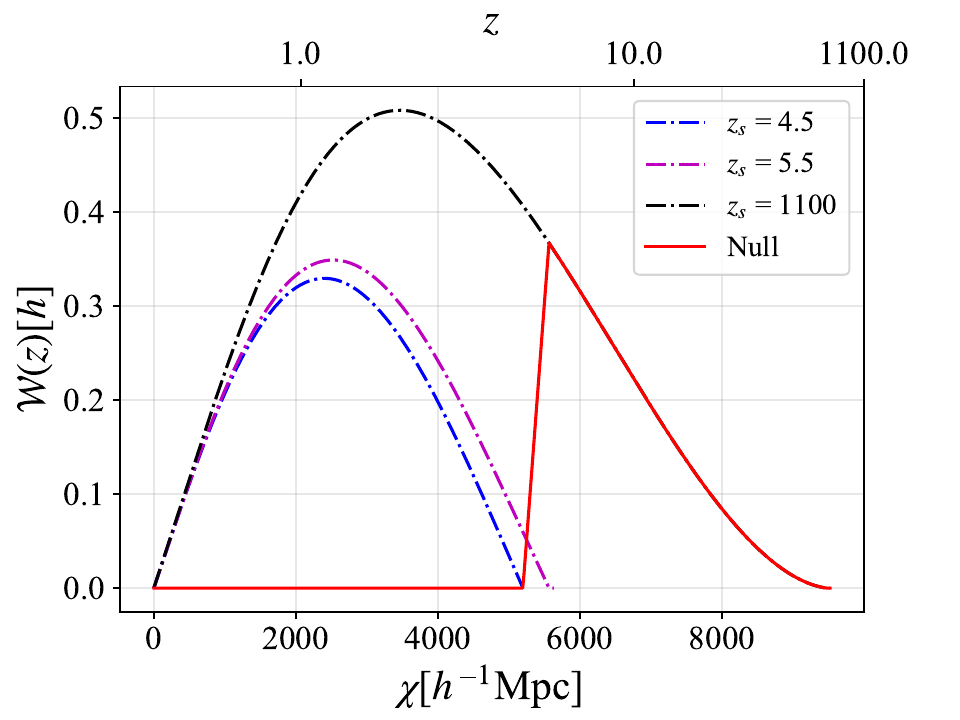}
\caption{The rescaled nulling kernel, defined as $\mathcal{W}(z,z_s) = W(z,z_s)\frac{c}{H(z)}$ for three sources. The CMB kernel is shown in black and the two LIM kernels at redshifts 4.5 and 5.5 are shown in blue and magenta, respectively. Finally, the LIM-nullling kernel is shown here in red and is null from $0<z<4.5$ which corresponds to a complete insensitivity to the matter density field over that redshift range.}
\label{fig:nulling}
\end{figure}

In Fig. \ref{fig:nulling}, the LIM lensing kernels, the CMB lensing kernel, and the LIM-nulling kernel are plotted. The LIM-nulling kernel is exactly null between $0<z<4.5$, meaning that when integrating over the whole redshift range in Eq. \eqref{eq:convergence}, $\kappa_{\rm{null}} = \kappa_{\rm{CMB}} + \alpha \kappa_{z_2} - (1+\alpha) \kappa_{z_1}
$ provides a map of the line of sight integrated matter density field between $4.5<z<1100$, providing a pristine view of early times. 

LIM-nulling can be thought of as a type of foreground cleaning where the LIM-lensing information is used to clean the low redshift contribution to the CMB lensing. The data product that results from LIM-nulling does not itself contain any LIM information, whether from the original map or from its lensing reconstruction. Of course, LIMs cannot partake in tracing out the matter density field at a time before the line emission was emitted. It is composed of CMB lensing information from $ z > z_{\rm{null}}$.

This high-redshift information can be captured statistically by computing the CMB $\times$ LIM-nulling convergence spectrum,
\begin{multline}\label{eq:CMBxNull_spectrum}
    C_{L}^{\hat{\kappa}\hat{\kappa}_{\rm{null}}} = \int_0^{z_s} \frac{W(z',z_{\rm{s}})W_{\rm{null}}(z',z_{\rm{1}},z_{\rm{2}},z_{\rm{CMB}})}{\chi(z')^2}
    \\ \times P_{\rm{m}}\left(k = \frac{L+1/2}{\chi(z')},z'\right)\frac{c \, dz'}{H(z')}.
\end{multline}
The motivation for always computing the cross-correlation of the LIM-nulling convergence with the CMB convergence and not simply in auto-correlation, is that cross-spectrum is free of all interloper bias, as shown in Ref. \cite{Maniyar_nulling}. In addition, the variance of the cross-spectrum contains fewer cross terms than the variance of the LIM-nulling auto-spectrum. This cross-spectrum, however, effectively contains the same cosmological information as the LIM-nulling auto-spectrum when $\Delta z$ is small. One therefore gains in signal-to-noise.

\begin{figure}[ht]
\includegraphics[width = 8.9cm]{ 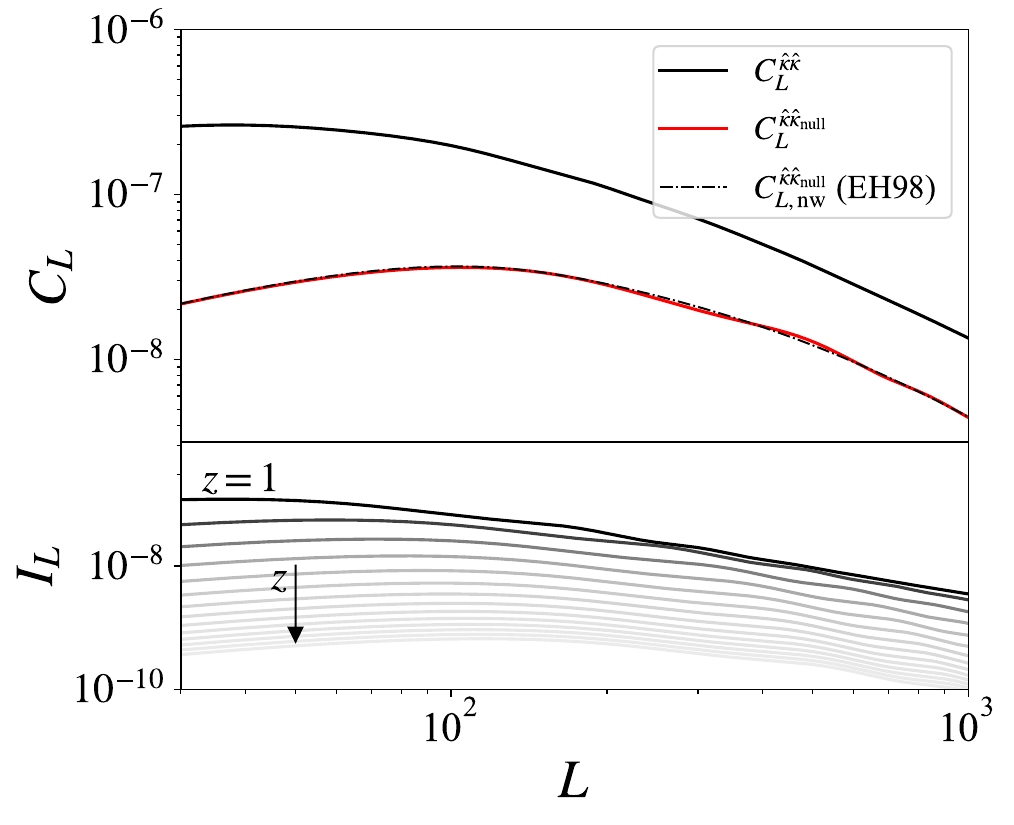}
\caption{Top: the CMB convergence spectrum, $C^{\hat{\kappa}\hat{\kappa}}_{L}$ in black and the CMB $\times$ LIM-nulling convergence spectrum, $C^{\hat{\kappa}\hat{\kappa}_{\rm{null}}}_{L}$, in red. The dot-dashed black curve shows the CMB $\times$ LIM-nulling convergence spectrum computed with the no-wiggle Eisenstein \& Hu fitting function. Bottom: the integrand of Eq. \eqref{eq:CMB_spectrum} evaluated at increasing redshifts from top to bottom starting at z = 1. Since the BAO scale is a fixed comoving scale, its angular projection changes as a function of $z$. The BAO features evolve gradually to lower $L$ as $z$ decreases which, when integrated over redshift, result in the washing out of BAO wiggles in $C^{\hat{\kappa}\hat{\kappa}}_{L}$. It is for this reason that the CMB convergence spectrum $C^{\hat{\kappa}\hat{\kappa}}_{L}$ is smooth with no discernible BAO features in the top panel. In contrast, the LIM-nulled convergence $C^{\hat{\kappa}\hat{\kappa}_{\rm{null}}}_{L}$ sees the reemergence of acoustic peaks (especially apparent when viewed against the reference no-wiggle nulled spectrum). These acoustic features are the result of the much slower angular evolution of BAO wiggles at early times.}
\label{fig:Cl_and_Il}
\end{figure}

In Fig. \ref{fig:Cl_and_Il} both the CMB convergence spectrum and the CMB $\times$ LIM-nulling convergence spectrum are plotted in black and red solid lines respectively. Upon first glance it is immediately evident that the CMB convergence has an order of magnitude more power than the CMB $\times$ LIM-nulling convergence spectrum. This is expected since a significant portion of the power has been nulled in the CMB $\times$ LIM-nulling spectrum. Perhaps more subtle is the re-emergence of acoustic peaks in the nulling spectrum. In order to help elucidate this feature,  we also show the CMB $\times$ LIM-nulling spectrum using the no-wiggle Eisentein and Hu fitting function in place of the typical matter power spectrum. This fitting function is essentially the matter power spectrum without the baryon acoustic oscillations (BAO). One can see the solid red line oscillating about the no-wiggle spectrum which is plotted in dashed black. We have refrained from forecasting the potential to constrain the BAO scale in this work as we believe it merits a dedicated discussion. In our companion paper, Ref. \cite{LIM_nulling_BAO}, we perform an Alcock-Paczynski test on  mock CMB $\times$ LIM-nulling data sets in order to forecast whether it is possible to measure BAO features with such a probe. 

\vspace{0.01in}

In the next two sections, we explore how the choice of various observing parameters affects the signal-to-noise ratio (SNR) of this probe. In order to do so, we require the variance of the CMB $\times$ LIM-Nulling cross-spectrum. It is given by
\begin{widetext}
\begin{eqnarray}\label{eq:nulling_var}
\textrm{var} (C^{\hat{\kappa}_{\rm{CMB}}\hat{\kappa}_{\rm{null}}}_{L}) &=& \left(\frac{1}{f_{\rm{sky}}(2L + 1)} \right) [2[ (C^{\hat{\kappa}_{\rm{CMB}}}_{L} + N^{\hat{\kappa}_{\rm{CMB}}}_{L})^2 + 2\alpha(C^{\hat{\kappa}_{\rm{CMB}}}_{L} + N^{\hat{\kappa}_{\rm{CMB}}}_{L})C^{\hat{\kappa}_{\rm{CMB}}\hat{\kappa}_{\rm{2}}}_{L} - 2(1+\alpha)(C^{\hat{\kappa}_{\rm{CMB}}}_{L} + N^{\hat{\kappa}_{\rm{CMB}}}_{L})C^{\hat{\kappa}_{\rm{CMB}}\hat{\kappa}_{\rm{1}}}_{L}   ] \nonumber \\&+&\alpha^2[(C^{\hat{\kappa}_{\rm{CMB}}\hat{\kappa}_{\rm{2}}}_{L})^2+(C^{\hat{\kappa}_{\rm{CMB}}}_{L} 
   + N^{\hat{\kappa}_{\rm{CMB}}}_{L})(C^{\hat{\kappa}_{\rm{2}}}_{L} + N^{\hat{\kappa}_{\rm{2}}}_{L})] -2\alpha(1+\alpha)[C^{\hat{\kappa}_{\rm{CMB}}\hat{\kappa}_{\rm{1}}}_{L}C^{\hat{\kappa}_{\rm{CMB}}\hat{\kappa}_{\rm{2}}}_{L}+ (C^{\hat{\kappa}_{\rm{CMB}}}_{L} + N^{\hat{\kappa}_{\rm{CMB}}}_{L} C_{L}^{\hat{\kappa}_{\rm{1}}\hat{\kappa}_{\rm{2}}}]\nonumber  \\
   &+&(1+\alpha)^2[ (C^{\hat{\kappa}_{\rm{CMB}}\hat{\kappa}_{\rm{1}}}_{L})^2+(C^{\hat{\kappa}_{\rm{CMB}}}_{L}
   + N^{\hat{\kappa}_{\rm{CMB}}}_{L})(C^{\hat{\kappa}_{\rm{1}}}_{L} + N^{\hat{\kappa}_{\rm{1}}}_{L})] ]\, ,
\end{eqnarray}
where we use ``$\rm{var}$"  to denote the variance. The quantity $N^{\hat{\kappa}_i}_{L}$ denote the lensing reconstruction noise corresponding to the estimated convergence $\hat{\kappa}_i$. To be explicit, the CMB reconstruction noise is denoted by $N^{\hat{\kappa}_{CMB}}_{L}$ and $N^{\hat{\kappa}_1}_{L}$ is shorthand for the LIM-pair lensing reconstruction noise,  $N_{XY}(L)$ , using lines X and Y at $z_{1}$. Finally, $f_{\rm{sky}}$ is the fraction of the sky area observed. A complete derivation can be found in Appendix \ref{app:nulling_var}.
\end{widetext}

\section{\label{sec:Choice_of_Lines} Cosmic Variance Limited SNR}

Before considering the effects of instrument noise on the CMB $\times$ LIM-nulling  SNR, we first wish to illustrate the importance of the choice of line redshifts and of the redshift separation between LIMs used for nulling. To do this, we work in harmonic space and compute the LIM-pair reconstruction noise at different redshifts as well as the CMB lensing reconstruction noise, both of which enter into Eq. \eqref{eq:nulling_var}. For various cases, we compute both the SNR per mode, as well as the cumulative SNR assuming uncorrelated errors which is given by 

\be
\label{eq:SNR}
\textrm{Cumulative SNR} = \left [\sum_{L}  \left( \frac{C_{L}}{\sqrt{\textrm{var} (C_{L})}}\right )^2 \right]^{1/2}.
\ee

We compute $C_{L}$ using the matter power spectrum from the publicly available code \texttt{CAMB}.\footnote{\url{https://github.com/cmbant/CAMB}} The line auto-spectra and cross-spectra which enter into the LIM-pair lenisng reconstruction noise through Eqs. \eqref{eq:F_XY} and \eqref{eq:lagrange_xy}, are obtained from the publicly available code \texttt{HaloGen}.\footnote{\url{https://github.com/EmmanuelSchaan/HaloGen/tree/LIM}} As the name suggests, \texttt{HaloGen} uses a halo model formalism based on conditional luminosity functions \cite{Schaan_2021_a,Schaan_2021_b}. For more information on how we model these lines here, we refer the interested reader to Appendix A of Ref. \cite{Maniyar_nulling}.
We take the spherically averaged power sepctra of each line and convert them into the corresponding angular spectra. Working in the thin shell approximation, this conversion is written as,

\begin{figure*}[ht]
\includegraphics[width=\textwidth,height=6cm]{ 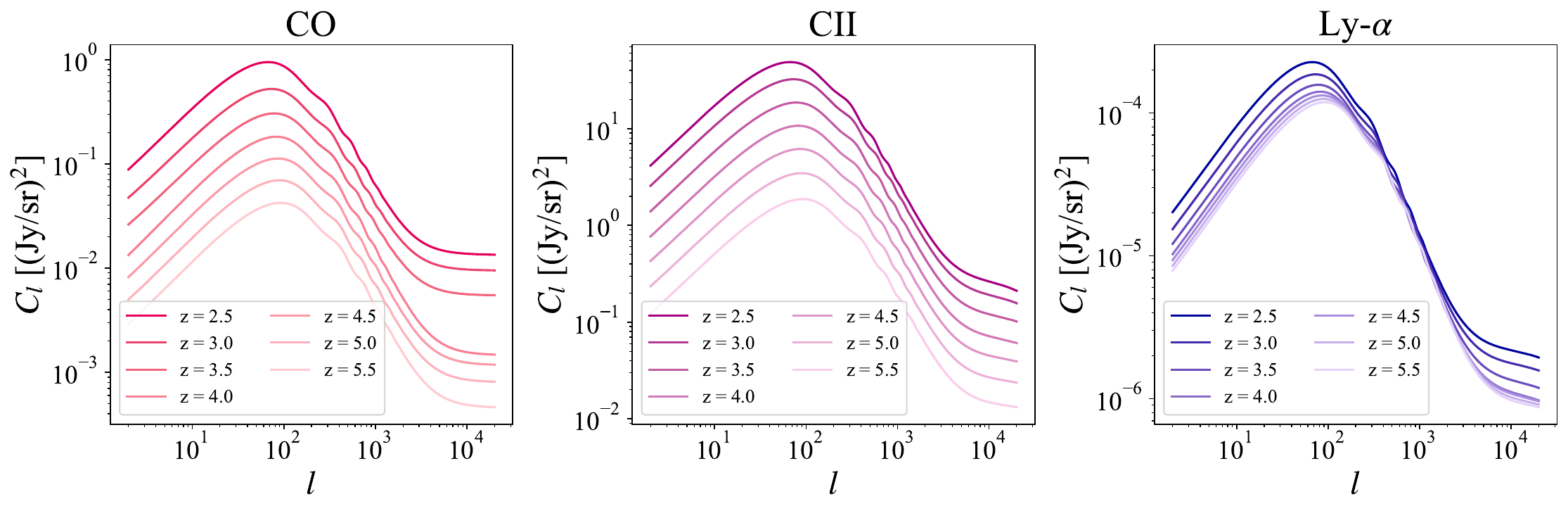}
\caption{Angular power spectra of CO 4-3 (left, red), [CII] (middle, purple), and Ly-$\alpha$ (right, blue) as functions of angular multipole $l$. We plot these spectra at redshifts ranging from $z = 2.5$ to $5.5$. Darker line colour denotes low-$z$ while paler line colour denoted high-$z$. To produce these spectra, spherically averaged line power spectra are obtained from \texttt{HaloGen} and then converted to angular power spectra using Eq. \eqref{eq:Pk_to_Cl}. Although the achromaticity of lensing (to first order) means that the amplitudes of the line power spectra cancel out in the noiseless cosmic variance-limited regime considered in Section~\ref{sec:Choice_of_Lines}, this changes in the realistic noisy scenarios that we examine in Section~\ref{sec:Sensitivity}.}
\label{fig:Line_Spectra}
\end{figure*}

\begin{figure}[ht]
\includegraphics[width = 8cm]{ 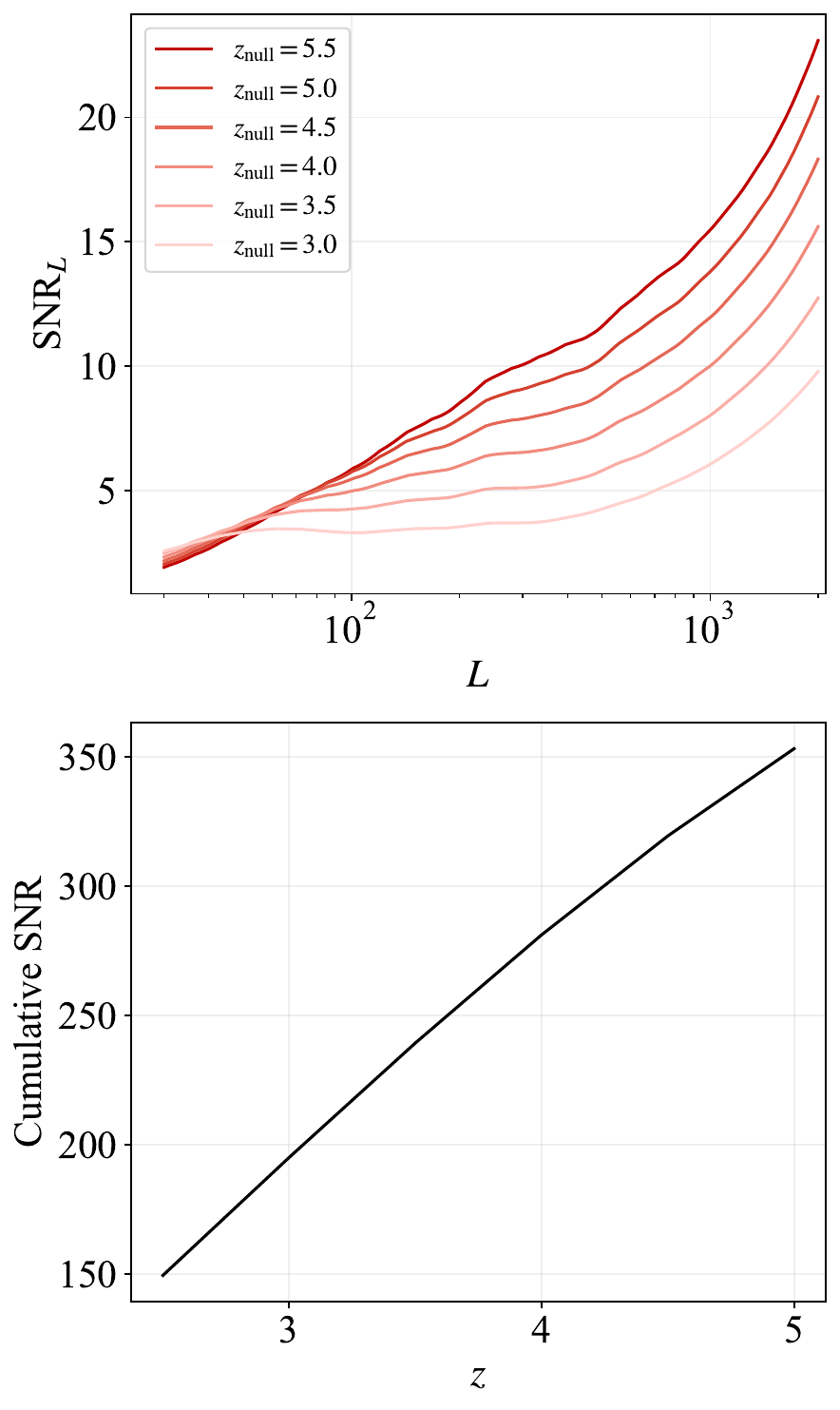}
\caption{ Top: signal-to-noise ratio of the nulling estimator per lensing multipole $L$. Each line denotes a different nulling redshift going from high redshift in dark red to low redshift in pale red. Bottom: The cumulative SNR as a function of nulling redshift. In all cases, the line separation is fixed at $\Delta z = 0.5$. Even aside from the science applications at high redshifts, one sees that there is preference for high-redshift nulling in order to maximize sensitivity.}
\label{fig:SNR_z}
\end{figure}

\begin{figure}[ht]
\includegraphics[width = 8cm]{ 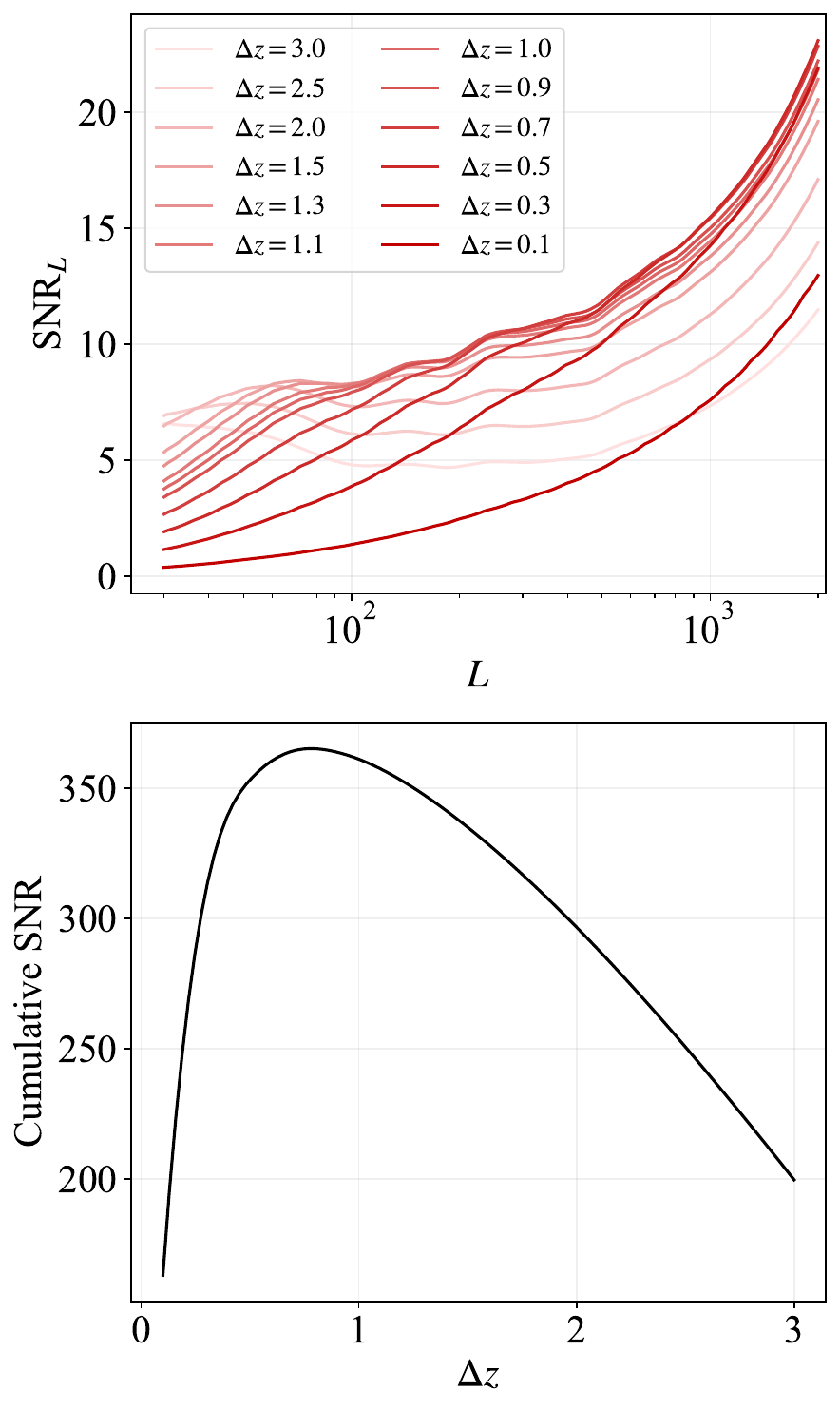}
\caption{Same as Fig.~\ref{fig:SNR_z}, but optimizing for the redshift separation $\Delta z$ between LIMs rather than the nulling redshift. In all cases, the higher redshift line is fixed at $z = 5.5$. In the noiseless cosmic variance-limited case the optimal SNR is achieved at $\Delta z = 0.7$, but this changes when instrumental noise is introduced in Section~\ref{sec:Sensitivity}.}
\label{fig:SNR_dz}
\end{figure}

\be\label{eq:Pk_to_Cl}
C_{l} = \mathcal{V}^{-1}P(k = l/\chi(z),z)
\ee
where $\mathcal{V} = \chi^2(z)\Delta \chi$ is the comoving volume per steradian of a shell centered at $z$. We set the width of the shell to the comoving distance corresponding to a redshift width of 0.05. 

We consider CO, [CII], and Ly-$\alpha$ as possible lines with which nulling could be performed. Their angular power spectra as a function of redshift are plotted in Fig. \ref{fig:Line_Spectra}. While the amplitude of the power spectra vary greatly between lines, devoid of noise and systematics, any pair of lines will yield the same SNR in the cosmic-variance-limited case; in this idealized case, the LIM-pair lensing reconstruction noise, $N_{XY}(L)$, is the same for any line pair, therefore the LIM-lensing SNR is fixed. Somewhat more intuitively, without the effects of noise and systematics, no line traces the line-of-sight gravitational potentials it encounters better or worse than any other since gravitational lensing is achromatic. This propagates through to nulling. Referring the reader back to Eq. \eqref{eq:CMBxNull_spectrum}, nowhere do the individual line spectra enter into $C_{L}^{\hat{\kappa}\hat{\kappa}_{\rm{null}}}$. 

The quantities of paramount importance, then, which do enter into both the CMB $\times$ LIM-nulling convergence as well as its variance, are the line redshifts. The redshifts, or comoving distance to the source plane, of the lines enters into the nulling kernel in two ways: through the LIM lensing kernels themselves and through the parameter $\alpha$. Referring back to Fig. \ref{fig:nulling} and \ref{fig:Cl_and_Il}, it might seem, by construction, that nulling at lower $z$ would increase the amplitude $C_{L}^{\hat{\kappa}\hat{\kappa}_{\rm{null}}}$, bringing it closer to $C_{L}^{\hat{\kappa}\hat{\kappa}}$, thus resulting in higher SNR. However, the line redshifts also enter into the nulling variance through the cross spectrum terms $C_{L}^{\hat{\kappa}_{i}\hat{\kappa}_{j}}$. This nulling variance decreases as a redshift increases. We find that these effects combined, the increase in both the nulling spectrum amplitude and the LIM-nulling variance as $z$ decreases, lead to a preference for high-$z$ nulling. Plotted in Fig. \ref{fig:SNR_z} is the CMB $\times$ LIM-nulling SNR as a function of $z$ and $L$, where the redshift separation of the lines is fixed at $\Delta z = 0.5$. As redshift increases the cumulative SNRs likewise increase.


The next choice to consider is the redshift separation of the lines. It would seem, looking at Fig. \ref{fig:nulling}, that an ideal nulling scneario would have the two lines be as close as possible in redshift in order to get a sharp cutoff. In Fig. \ref{fig:SNR_dz}, the SNR as a function of redshift separation, $\Delta z$, is plotted. It is immediately obvious that both very small and very large separations of the lines are sub-optimal. The cumulative SNR peaks at $\Delta z = 0.7$ when nulling performed with the higher redshift LIM at z = 5.5.

It is important to note that $\Delta z = 0.7$ is not the optimal solution in general, since what must be optimized in the nulling equation is the interplay between the comoving distance separation of the three probes being used. While throughout this work we consider nulling performed on the CMB lensing convergence by LIMs, any three convergence maps may be used for nulling. For instance galaxy lensing convergence maps from $z\sim 2$, can be used to null a LIM convergence map at $z = 10$. The redshift optimization must be performed on a case-by-case basis since what is at work here is the interplay between the comoving distances of the three probes used in the nulling estimator. What is more, the addition of interlopers, diffuse foreground contaminants, and instrument systematics, which have non-trivial frequency evolution, complicate the problem. Care must be taken to optimize for the observing scenario at hand. 

In this section, we provide the reader with intuition about how the LIM-nulling estimator depends on LIM parameters. We show that in the idealized case where noise and systematics are not included, the SNR depends only on the choice of line redshift and the redshift separation of the lines. Illustrating this with a concrete example, we presented the optimal redshift solutions for CMB nulling taking place at $z = 4.8$ in the cosmic variance limited case. We found that the LIM redshift which maximize the cumulative SNR are $z_{1} = 4.8$ and $z_{2} = 5.5$. In the presence of the frequency dependent noise of LIM experiments, this breaks down. In the following section, we explore how the CMB $\times$ LIM-nulling SNR scales with the sensitivity and the area of mm, sub-mm and IR surveys. We choose three fiducial observing scenarios to use in the subsequent forecasts. These include instrument noise and line interlopers, and we therefore re-optimize the LIM redshifts given these contaminants.

\section{\label{sec:Sensitivity} Survey Area and Sensitivity}
In this section we explore the dependence of survey specifications on the CMB $\times$ LIM-nulling SNR. As previously mentioned, when devoid of instrument noise and systematics, all lines yield the same lensing reconstruction noise and therefore the same nulling SNR. Once such effects are added, this is no longer the case since the map-level SNR enters into $N_{L}^{\hat{\kappa}}$. In Section \ref{subsec:Instrument_Noise},  we present the models used for computing the noise power of CO, [CII], and Ly-$\alpha$-type experiments. In Section \ref{subsec:Sensitivity} we study how our nulling statistic depends on the intensity mapping survey sensitivities, while in Section \ref{subsec:Survey_Area}, we explore how the SNR depends on their survey areas. Subsequently, we optimize the survey area for a given survey sensitivity to maximize the nulling SNR.  

While the final forecasts in Sections \ref{sec:LCDM Forecast} and \ref{sec:PCA} make use of the LIM-pair estimator, this exploratory section assumes each LIM lensing convergence map is estimated using a single line at a time. This is done to isolate the effects of a particular instrument in order to explore how the CMB $\times$ LIM-nulling SNR varies as a function of survey area and sensitivity for a single instrument. For example, we use [CII] observations from two channels of a given [CII] survey to perform nulling. 

\subsection{\label{subsec:Instrument_Noise} Instrument Noise Power}

\subsubsection{CO Experiments}
In the case of CO experiments, we typically write the noise power as 
\be\label{eq:Pn_CO}
P^{\rm{CO}}_N=\sigma_{\rm{vox}}^2V^{\rm{CO}}_{\rm{vox}},
\ee
where $\sigma_{\rm{vox}}$ is the noise in a single voxel and is given by 
\be\label{eq:sigma_CO}
\sigma_{\rm{vox}}=\frac{T_{\rm{sys}}}{\sqrt{N_{\rm{det}}t^{\rm{CO}}_{\rm{pix}}\delta\nu }}.
\ee
Here, $T_{\rm{sys}}$ is the system temperature, $N_{\rm{det}}$ is the number of detector feeds, and $t^{\rm{CO}}_{\rm{pix}}$ is the observing time of a single pixel \cite{Li_2016}. The time per pixel is related to the total observing time $t_{\rm{obs}}$ of the survey via $t^{\rm{CO}}_{\rm{pix}}= t_{\rm{obs}}(\Omega^{\rm{CO}}_{\rm{pix}}/\Omega_{\rm{surv}})$  where $\Omega_{\rm{surv}}$ is the total survey area. A ``pixel" here is defined to cover a solid angle $\Omega^{\rm{CO}}_{\rm{pix}}=\sigma^2_{\rm{beam}}$,  where $\sigma^{2}_{\rm{beam}}$ is the variance of the instrument's Gaussian beam. We compute $\sigma_{\rm{beam}}$ using the relation $\sigma_{\rm{beam}} = \theta_{\rm{FWHM}}/\sqrt{8 \ln 2}$ where $\theta_{\rm{FWHM}}$ is the beam full width half maximum. The comoving volume $V^{\rm{CO}}_{\rm{vox}}$ of a single voxel is the volume subtended by a pixel with angular size $\Omega^{\rm{CO}}_{\rm{pix}}$ and frequency resolution $\delta\nu$.   Since the number of detectors, $N_{\rm{det}}$, constitutes a measure of the instantaneous sensitivity of the instrument, we parameterize the total survey sensitivity in terms of spectrometer-hours, $N_{\rm{det}}t_{\rm{obs}}$.

In the following subsection, we demonstrate how the SNR of the CMB $\times$ LIM-nulling convergence spectrum scales with survey area (or equivalently, $f_{\rm{sky}}$, the fractional sky coverage) and sensitivity, $N_{\rm{det}}t_{\rm{obs}}$. To that end, we compute the CO noise power using Eq. \eqref{eq:Pn_CO} and include it in our computation of the LIM lensing reconstruction noise, the single line analog to the LIM-par lensing estimator in Eq. \eqref{eq:LIM-pair estimator}.  When varying the sensitivity and survey area, we anchor the remaining instrument specifications to the CO Mapping Array Project phase 2 (COMAP2) \cite{Ihle_2019}. The specifications of this instrument as well as its current generation phase 1 counterpart (COMAP1) can be found in Table \ref{tab:CO_experiments}. COMAP aims to detect spectral lines from various rotational line transitions of CO, including the CO(1-0) during the peak of star formation around $z \sim 3$ when the CO luminosity function peaks, and will have some sensitivity of other CO transition lines out to $z\sim 8$ \cite{Cleary_2022,Breysse_2022,Chung_2022}.

\begin{table}[b]
\centering
\caption{Instrument parameters for COMAP1 and COMAP2 experiments. These values are taken from Ref. \cite{Ihle_2019}.}
\setlength{\tabcolsep}{15pt} 
\begin{tabular}{lcc}\toprule
Parameter                                                          & COMAP1 & COMAP2~ \\\midrule
$T_{\rm{sys}}$ (K)                                & 40      & 40        \\
$N_{\rm{det}}$                              & 19      & 95        \\
$\theta_{\rm{FWHM}}$ (arcsec)   & 4       & 4         \\
$\Delta\nu$ (MHz)            & 15.6       & 15.6         \\
$t_{\rm{obs}}$ (h)                   & 6000    & 9000      \\
$\Omega_{\rm{surv}}$ (deg$^2$) & 2.5       & 2.5        \\ \bottomrule
\end{tabular}

\label{tab:CO_experiments}
\end{table}

\subsubsection{[CII] Experiments}

In the case of [CII] surveys, we follow Ref. \cite{Chung_2018} and write
\begin{equation}\label{eq:Pn_CII}
P^{\rm{CII}}_N=\frac{\sigma_{\rm{pix}}^2}{t_{\rm{pix}}^{\rm{CII}}}V_{\rm{vox}}^{\rm{CII}}.
\end{equation}
Here, $t_{\rm{pix}}^{\rm{CII}}$ and $V_{\rm{vox}}^{\rm{CII}}$ are defined (by convention) slightly differently than the corresponding CO quantities. A ``pixel" is defined to cover a solid angle  $\Omega_{\rm{pix}}^{\rm{CII}}=2\pi\sigma_{\rm{beam}}^2$.  This new definition of the pixel size is then used, along with the frequency channel width, to compute $V_{\rm{vox}}^{\rm{CII}}$. Here, 
\begin{equation}\label{eq:sigma_CII}
t_{\rm{pix}}^{\rm{CII}}=\frac{N_{\rm{det}}t_{\rm{obs}}}{\Omega_{\rm{surv}}/\Omega_{\rm{pix}}^{\rm{CII}}}. 
\end{equation}

Analogously to the previous section on CO, when varying the sensitivity and survey area we fix the rest of the instrument parameters to those of the the CarbON CII line in post-rEionisation and ReionisaTiOn epoch (CONCERTO), a current generation high-$z$ [CII] mapping experiment \cite{CONCERTO}. CONCERTO will detect the [CII] line from $6 \lesssim z \lesssim 11$ over 1.4 deg$^{2}$ on the sky.  The instrument specification used for simulating CONCERTO noise are summarized in Table \ref{tab:CII_experiments}. We also compute the SNR for a handful of other [CII] mapping experiments whose specifications are also summarized in Table \ref{tab:CII_experiments}.

\begin{table}
\centering
\caption{Instrument parameters for [CII] mapping experiments, including the Fred Young Submillimeter Telescope (FYST), CONCERTO, Tomographic Ionized Carbon Intensity Mapping Experiment (TIME), and a next-generation ``Stage II" concept. The Stage II parameters are
based on Ref. \cite{Silva_2015}, while the rest are from Ref. \cite{Chung_2018}.}
\setlength{\tabcolsep}{4pt} 
\begin{tabular}{lcccc}\toprule
Parameter & FYST & CONCERTO & TIME & StageII \\
\midrule
$\sigma_{\rm{pix}}$ (MJy/sr s$^{1/2}$) & 0.86 & 11.0 & 11.0 & 0.21 \\
$N_{\rm{det}}$ & 20 & 3000 & 32 & 16000 \\
$\theta_{\rm{FWHM}}$ (arcsec) & 46.0 & 22.5 & 22.5 & 30.1 \\
$\delta\nu$ (GHz) & 2.5 & 1.5 & 1.9 & 0.4 \\
$t_{\rm{obs}}$ (hr) & 4000 & 1200 & 1000 & 2000 \\
$\Omega_{\rm{surv}}$ (deg$^2$) & 16 & 1.4 & 0.6 & 100 \\
\bottomrule
\end{tabular}
\label{tab:CII_experiments}
\end{table}

\subsubsection{Ly-\texorpdfstring{$\alpha$}{TEXT} Experiments}

For the noise power associated with infrared (IR) intensity mapping experiments targeting the high-$z$ Ly-$\alpha$ line, we write the noise power spectrum $P^{\textrm{Ly-}\alpha}_{N}$ as  
\be\label{eq:Pn_lya}
P^{\textrm{Ly-}\alpha}_{N} =  \sigma_{\rm{vox}}^2 V^{\textrm{Ly-}\alpha}_{\rm{vox}}
\ee
where $V^{\textrm{Ly-}\alpha}_{\rm{vox}}$ is the single-voxel volume (defined in an identical way to analogous quantities as the [CII] and CO cases) and $\sigma_{\rm{vox}}$ is defined as
\be\label{eq:sigma_lya}
\sigma_{\rm{vox}} = s \sqrt{\frac{4\pi f_{\rm{sky}}}{t_{\rm{obs}}\Omega_{\rm{pix}}}}.
\ee
Here the instantaneous pixel sensitivity is given by $s$. Since the IR experiments we consider scan the sky one pixel at a time, the number of detectors is simply unity and we thus parameterize our total survey sensitivity by the total observing time, $t_{\rm{obs}}$, instead of spectrometer-hours. We set the other instrument specifications in Eqs. \eqref{eq:Pn_lya} and \eqref{eq:sigma_lya} to that of the Cosmic Dawn Intensity Mapper (CDIM) \cite{CDIM}. CDIM is a next generation optical and IR instrument aimed at detecting high redshift galaxies and quasars as well as spectral lines during cosmic dawn and reionization. We also compute the SNR for Spectro-Photometer for the History of the Universe, Epoch of Reionization and Ices Explorer (SPHEREx), an upcoming intensity mapping mission with Ly-$\alpha$ mapping capabilities \cite{SPHEREx}. The specification for both CDIM and SPHEREx are summarized in Table \ref{tab:IR_experiments}.

\begin{table}
\centering
\caption{Instrument parameters for Ly-$\alpha$ experiments. Instrument specification for SPHEREx and CDIM are obtained from Ref. \cite{SPHEREx} and Ref. \cite{CDIM} respectively. We compute the sensitivity, $s$, for a nominal 2000 hour survey given the quoted noise powers for both instruments. For SPHEREx, the spectral resolving power ($R\equiv \lambda / \Delta \lambda$, where $\lambda$ is the observing wavelength and $\Delta \lambda$ is the wavelength resolution) is computed at $\lambda \simeq 4.5\,\mu m$. CDIM achieves R $\geq 300$ over its whole bandwidth.}
\setlength{\tabcolsep}{19pt} 
\begin{tabular}{lcc}\toprule
Parameter                                                          & SPHEREx & CDIM \\\midrule
s (Jy/sr s$^{1/2}$)                               & 231.46     & 38.58        \\
$\theta_{\rm{FWHM}}$ (arcsec)   & 6       & 2         \\
$t_{\rm{obs}}$ (h)                   & 2000    & 2000      \\
$\Omega_{\rm{surv}}$ (deg$^2$) & 100       & 100         \\ 
$R \equiv \lambda/\Delta\lambda$ & 150       & 300        \\\bottomrule
\end{tabular}
\label{tab:IR_experiments}
\end{table}

\subsection{\label{subsec:Sensitivity} Dependence on Sensitivity}

We compute the CMB $\times$ LIM-nulling convergence spectrum and its variance when nulling is performed with LIMs from $z = 5.5$ and $z = 3.5$. The LIM lensing reconstruction noise is computed using the single line analog of Eq. \eqref{eq:limpair_variance}. We take $l_{\rm{min,LIM}} = 30$ and $l_{\rm{max,LIM}} = 5000$ where $l_{\rm{min}}$ is driven by the area of the survey and $l_{\rm{max}}$ is driven by the angular resolution of the instrument.For the CMB lensing reconstruction noise, we assume that of SO and use $N^{\hat{\kappa}}_{L}$ from the SO noise calculator.\footnote{\url{https://github.com/simonsobs/so\_noise\_models/tree/master/LAT\_lensing\_noise}}   It may be noted that some of the instruments we consider in this section do not probe modes as large as $l = 30$. For the purposes of singling out the effect of increasing sensitivity, we fix $l_{\rm{min,LIM}} = 30$  when varying the sensitivity. When computing the LIM-nulling SNR of particular experiments in following sections, we adjust $l_{\rm{min,LIM}}$ accordingly. 

\begin{figure*}[ht]
\includegraphics[width=\textwidth]{ 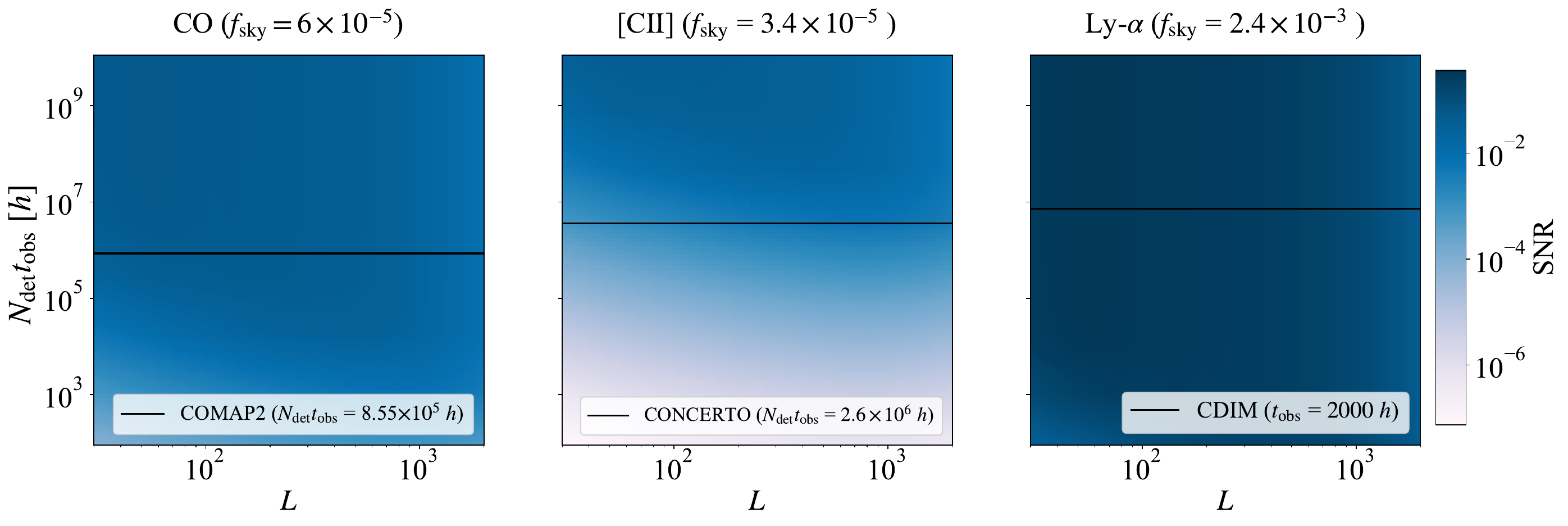}
\caption{SNR as a function of lensing multipole, $L$, and sensitivity in terms of spectrometer-hours, $N_{\rm{det}}t_{\rm{obs}}$, for CO-type experiments (left), [CII]-type experiments (middle), and Ly-$\alpha$-type experiments (right). For Ly-$\alpha$, since $N_{\rm{det}} = 1$, the sensitivity is simply parameterized by $t_{\rm{obs}}$. The black horizontal line in each panel denotes the sensitivity of each line's nominal survey.}
\label{fig:Sensitivity_Lz}
\end{figure*}

\begin{figure}[ht]
\includegraphics[width=8.6cm]{ 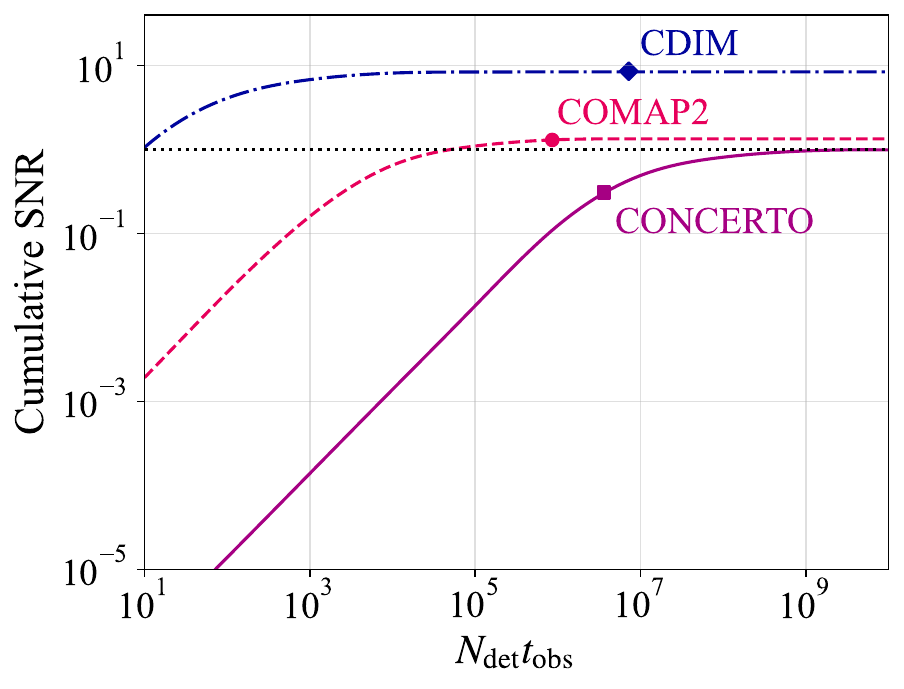}
\caption{Cumulative SNR as a function of $N_{\rm{det}}t_{\rm{obs}}$ for Ly-$\alpha$-type experiments (blue dot-dashed), CO-type experiments (pink dashed), and [CII]-type experiments (solid purple). The dotted black line denotes SNR = 1. It should be noted that for Ly-$\alpha$ experiments, $N_{\rm{det}} = 1$ and so the horizontal axis can simply be interpreted as $t_{\rm{obs}}$. The cumulative SNR for the nominal CDIM (blue diamond), COMAP2 (pink circle), and CONCERTO (purple square) surveys are also shown here. With the (mild) exception of CONCERTO, all three experiments roughly sit on the plateau where further increases in the number of spectrometer-hours do not increase SNR. }
\label{fig:Sensitivity_cumulative}
\end{figure}

In Fig. \ref{fig:Sensitivity_Lz}, we plot the SNR per $L$ mode as a function of $L$ and spectrometer-hours $N_{\rm det} t_{\rm obs}$. Summing over all $L$, the {\it cumulative} SNR as a function of the number of spectrometer hours is plotted in Fig.~\ref{fig:Sensitivity_cumulative}. As expected, cumulative SNR increases as  $N_{\rm{det}}t_{\rm{obs}}$ increases until it saturates to a plateau.
For COMAP2-type instruments, this plateau at  $N_{\rm{det}}t_{\rm{obs}} \sim 10^4\,\textrm{h}$, while for CONCERTO- and CDIM-type experiments this occurs at $N_{\rm{det}}t_{\rm{obs}} \sim 10^6\,\textrm{h}$ and $N_{\rm{det}}t_{\rm{obs}} \sim 10^2\,\textrm{h}$ respectively .

Using the nominal survey specifications listed in Table \ref{tab:CO_experiments}, COMAP1 yields a cumulative SNR of 1.11 while COMAP2 has a cumulative SNR of 1.30. As for how current and upcoming [CII] experiments fare, Stage II has a cumulative SNR of 8.33 while all other experiments yield a cumulative SNR $< 1$. Nulling with SPHEREx results in a noise dominated measurement; however, CDIM achieves a cumulative SNR of 7.05. For CDIM, COMAP2, and CONCERTO, the nominal survey configurations are denoted with the diamond, circle, and square in Fig.~\ref{fig:Sensitivity_cumulative}, respectively. It is clear that perhaps with the exception of CONCERTO, these configurations are in the regime where the cumulative SNR has plateaued as a function of spectrometer-hours. Therefore, in order to increase the SNR further, other experimental parameters must be altered. 

\subsection{\label{subsec:Survey_Area} Dependence on Survey Area}

Next we consider how the CMB $\times$ LIM-nulling SNR varies as a function of the survey area, parametrized by the fractional sky area $f_{\rm{sky}} \equiv \Omega_{\rm{surv}}/4\pi$. For any cosmological measurement there is always a trade-off when increasing $f_{\rm{sky}}$. Keeping the sensitivity and observing time fixed, increasing the sky coverage of a survey means that less time is spent integrating on each pixel. This results in a shallower survey. This can be seen by examining Eqs. \eqref{eq:Pn_CO} and \eqref{eq:sigma_CO}, which show that the noise power is proportional to $f_{\rm{sky}}$. However, increasing the sky coverage also increases the number of Fourier modes sampled and results in decreased sample variance per mode. This is reflected in Eq. \eqref{eq:nulling_var} where the nulling variance sees a factor of $ 1/ f_{\rm{sky}}$ out front. It is the optimization of these two effects that determines the optimal survey coverage for an instrument of a given sensitivity \cite{Tegmark_noise_1997}. 

\begin{figure*}[ht]
\includegraphics[width=\textwidth]{ 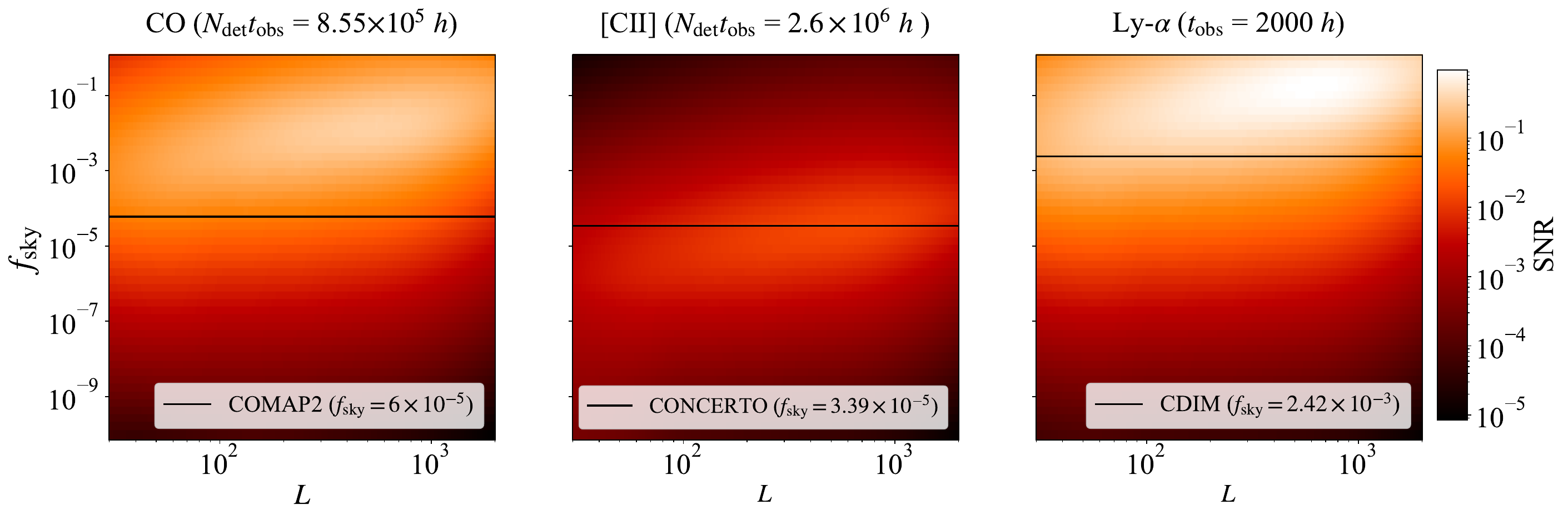}
\caption{SNR as a function of lensing multipole, $L$, and survey area, $f_{\rm{sky}}$, for CO-type experiments (left), [CII]-type experiments (middle), and Ly-$\alpha$-type experiments (right). The black horizontal line in each panel denotes the survey area of each line's nominal survey.}
\label{fig:Fsky_Lz}
\end{figure*}

\begin{figure}[ht]
\includegraphics[width=8.6cm]{ 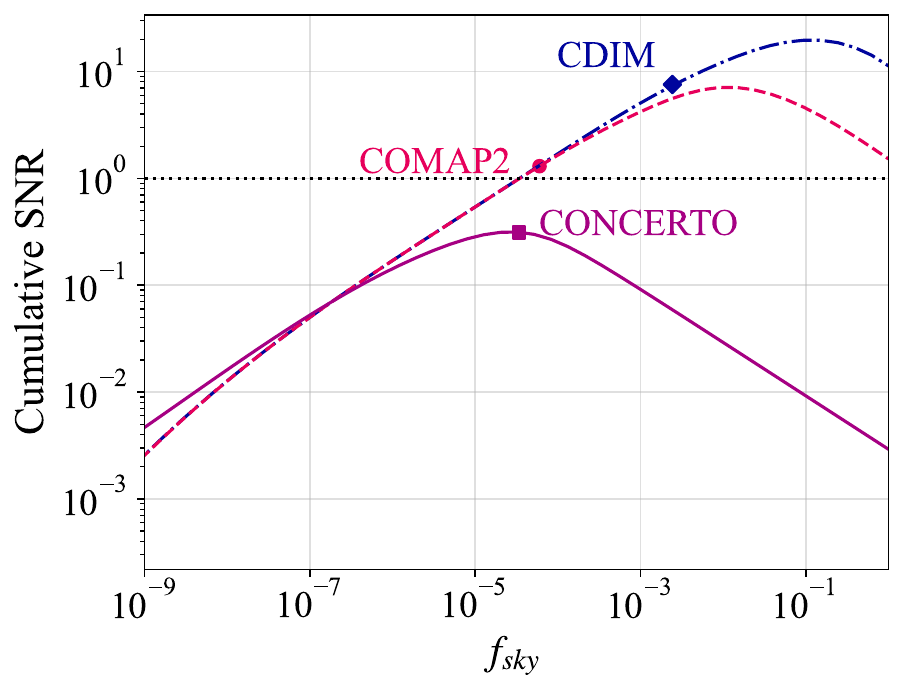}
\caption{ Cumulative SNR as a function of $f_{\rm{sky}}$ for Ly-$\alpha$-type experiments (blue dot-dashed), CO-type experiments (pink dashed), and [CII]-type experiments (solid purple). The dotted black line denotes SNR = 1. The cumulative SNR for the nominal CDIM (blue diamond), COMAP2 (pink circle), and CONCERTO (purple square) surveys are also shown here. While CONCERTO is reasonably optimized to balance instrumental noise and cosmic variance, CDIM and COMAP2 would benefit from additional sky coverage as far as a nulling measurement is concerned.}
\label{fig:Fsky_cumulative}
\end{figure}

In Fig.~\ref{fig:Fsky_Lz} we plot the SNR as a function of $L$ and $f_{\rm{fsky}}$ while in Fig.~\ref{fig:Fsky_cumulative} , we simply plot the cumulative SNR as a function of $f_{\rm{sky}}$. We compute $L$ up to 2000. This somewhat arbitrary cutoff was informed by our reconstruction noise curves which increase at high $L$. In addition, the small angular scales at the map level used in the reconstruction contain non-linear baryonic effects which are difficult to model accurately. For all three types of instruments, the cumulative SNR curves follow the same shape. The slope increases until it reaches a maximum, and subsequently decreases. This maximum survey area balances the two competing effects: the survey depth and the sample variance per mode. If the survey area is too small for a fixed survey duration, not enough modes are sampled and the SNR decreases. The region of the cumulative SNR curves to the left of the maximum value constitute this regime. If the survey area is too big for a fixed survey duration, many modes are sampled, but the survey is shallow, thus decreasing the cumulative SNR. This constitutes the region to the right of the maximum value. 

CONCERTO is an excellent example of a survey whose area yields the maximum CMB $\times$ LIM-nulling SNR given its sensitivity. This is not by chance. The 1.4 deg$^2$ survey area for CONCERTO optimizes the SNR of its line power spectrum and we find that as an approximate rule of thumb, optimizing the line power spectrum SNR with respect to $f_{\rm{sky}}$ also maximizes the nulling SNR.

In contrast, we see from Fig.~\ref{fig:Fsky_cumulative} that for COMAP2 and CDIM, the survey areas are not optimal. At present, the sky coverage of these surveys place them in the sample variance limited regime where the instantaneous sensitivity is relatively high, but simply not enough modes are sampled to truly maximize the capabilities of the instrument. 

\subsection{\label{subsec:Forecast_Scenarios}  Forecast Scenarios}

Given the exploration into the observational parameter space of the last few sections, we now present the three observing scenarios that are used in our cosmological parameter estimation forecasts. Here we return to computing the full CMB $\times$ LIM-nulling convergence using the LIM-{\it pair} estimator, where the convergence maps at each redshift are constructed with two different lines via Eq.~\eqref{eq:LIM-pair estimator}. We denote the scenarios as Current Generation, Next Generation, and Futuristic. We choose to perform LIM-nulling with LIM observations from [CII] and Ly-$\alpha$ mapping experiments given that these lines yielded higher SNRs in the near term. These scenarios are summarized in Table \ref{tab:Scenarios}. 

For our Current Generation scenario, we take [CII] to be observed by CONCERTO at $z = 5.5$ and Ly-$\alpha$ to be observed by SPHEREx at $z = 3.5$. In Section~\ref{sec:Choice_of_Lines}, we optimized the line separation in the noiseless case to be $\Delta z = 0.7$. In the presence of frequency-dependent noise, that is no longer the optimal line separation. Given the noise power of CONCERTO and SPHEREx, the line redshifts which maximize the SNR are $z = 5.5$ and $z = 3.5$. Of course, aside from pure sensitivity concerns, one must also account for systematics such as interloper lines. At high redshifts, Ly-$\alpha$ is contaminated by low-redshift H-$\alpha$ emission while the high-redshift [CII] line is contaminated by low-redshift CO emission. Using \texttt{HaloGen}, we generate H-$\alpha$ and CO spectra at the appropriate redshifts and include these when computing the LIM-pair lensing reconstruction noise. We assume the observations have undergone foreground removal, leaving behind a 10\% interloper residual power for each line. To simulate this, we simply multiply the generated interloper spectra by an overall factor of 0.1.

Although SPHEREx has two large 100 deg$^{2}$ deep fields at the poles, CONCERTO has a much smaller field of view, at 1.4 deg$^{2}$. Therefore, $\kappa_{\rm{null}}$ can only be computed over the small overlapping field of 1.4 deg$^{2}$. As a result, we compute LIM-lensing reconstruction noise with the largest angular scale $l_{\rm{min,LIM}} = 153$ ($\sim\!\!1\,{\rm deg}$). We take the finest angular scale to be $l_{\rm{max,LIM}} = 5000$ ($\sim\!0.04\,{\rm deg}$), which is much coarser than the angular resolution of both instruments. For the CMB instrument we choose SO where lensing reconstruction is performed with minimum and maximum spherical harmonic $l$ of $l_{\rm{min,CMB}} = 30$ and $l_{\rm{max,CMB}} = 5000$, respectively. The total SNR in this scenario is 0.1, suggesting that nulling estimation will likely be a future endeavour.

For the Next Generation scenario, we take [CII] to be observed by a Stage II instrument at $z = 5.5$ and Ly-$\alpha$ to be observed by CDIM at $z = 4.5$. Again, these redshifts are obtained by optimizing the nulling SNR for a given noise level. We assume the observations contain 5\% interloper residual power for each line. Here, both instruments are expected to survey a $100\,{\rm deg}^{2}$ field and we assume that they overlap entirely. We compute the LIM-lensing reconstruction noise with $l_{\rm{min,LIM}} = 30$ ($\sim\!6\,{\rm deg}$) and $l_{\rm{max,LIM}} = 10000$ ($\sim \! 0.02\,{\rm deg}$), again, using scales coarser  than to angular resolution of both instruments. For the CMB instrument we simulate the noise power of CMBS4 where lensing reconstruction is performed with  $l_{\rm{min,CMB}} = 30$ and $l_{\rm{max,CMB}} = 5000$. The CMBS4 lensing reconstruction noise used here can be found in Ref. \cite{Sailer_2021}. The total SNR for the Next Generation scenario is 9.5, representing a firm detection that will be an important proof-of-concept for the nulling technique. However, it will perhaps still not quite be the high-precision measurement that unlocks high-precision science.

Finally, we construct a Futuristic scenario that guarantees a high SNR measurement. We consider [CII] observed by a Stage II-like instrument at $z = 5.5$ and Ly-$\alpha$ to be observed by a CDIM-like instrument at $z = 4.5$, over a quarter of the sky. By this we mean that the [CII] maps are the same depth as those expected from Stage II but over a larger area of the sky. Likewise the CDIM-like instrument is one that produces maps at the same depth as the nominal CDIM survey but again over a larger portion of the sky. This can be achieved by using an instrument with the same instantaneous sensitivity and increasing the total observing time of the survey until the instantaneous integration time reaches that of the nominal surveys. This can also be achieved by increasing the scanning rate of the instrument such as to observe a larger portion of the sky in the same total observing time, but correspondingly increasing the instantaneous sensitivity in order to obtain the same depth. Of course, a combination of these two strategies will also suffice. We assume the observations contain 1\% interloper residual power for each line. Like the last scenario, we compute the LIM-lensing reconstruction noise with $l_{\rm{min,LIM}} = 30$  and $l_{\rm{max,LIM}} = 10000$. For the CMB instrument we again choose the CMBS4 where lensing reconstruction is performed with  $l_{\rm{min,CMB}} = 30$ and $l_{\rm{max,CMB}} = 5000$. The Futuristic scenario has a total SNR of $110$. As we will demonstrate in Section~\ref{sec:LCDM Forecast}, in this regime one is able to obtain competitive parameter constraints.

In Table \ref{tab:Scenarios}, we also quote the total SNR for a cosmic variance limited case to showcase the upper bound of what is achievable in an idealized case. Here we include no instrument noise nor interloper contaminants. Like the Futuristic scenario, we assume $l_{\rm{min,LIM}} = 30$  and $l_{\rm{max,LIM}} = 10000$ and $l_{\rm{min,CMB}} = 30$ and $l_{\rm{max,CMB}} = 5000$, but we set $f_{\rm{sky}} = 1$. In this case, the CMB $\times$ LIM-nulling cumulative SNR is 408.

\begin{table*}[ht]
\centering
\setlength{\tabcolsep}{4pt} 
\renewcommand{\arraystretch}{1.3} 
\begin{tabular}{lcccccccc}\toprule
Scenario          & Ly-$\alpha$ & [CII] & CMB & Survey Area         & $l_{\rm{min,LIM}}$ & $l_{\rm{max,LIM}}$ & Interloper Residual & $\sum$SNR \\\midrule
Current Generation & SPHEREx                    & CONCERTO            & SO                 & 1.4 deg$^2$         & 153              & 5,000            & 10\%                     & 0.1            \\
Next Generation    & CDIM                       & Stage II            & CMBS4              & 100 deg$^2$         & 30               & 10,000           & 5\%                      & 9.5            \\
Futuristic         & CDIM                       & Stage II            & CMBS4              & $f_{\rm{sky}} = 0.25$ & 30               & 20,000           & 1\%                      & 110            \\
CV Limited         & --                          & --                   & --                  & $f_{\rm{sky}} = 1$    & 30               & 20,000           & --                        & 408       \\\bottomrule      
\end{tabular}
\caption{Observational scenarios used for our forecasts, in addition to the cosmic variance limited case as a high-SNR reference. While current instruments will not make nulling detections, their successors will be capable of not just detections but also high-SNR characterizations that will be scientifically interesting.}\label{tab:Scenarios}
\end{table*}

\section{\label{sec:LCDM Forecast} Fisher Forecast: \texorpdfstring{$\Lambda$}{TEXT}CDM + \texorpdfstring{$M_{\nu}$}{TEXT} Cosmology}

In this section we present a Fisher forecast on potential parameter constraints from CMB $\times$ LIM-nulling when $0 < z \lesssim 5$ has been nulled. We take our parameter set to be the standard $\Lambda$CDM parameters plus the sum of the neutrino masses $M_{\nu}$. We begin by providing a brief overview of the Fisher formalism to establish notation, and follow by presenting the parameter covariances for our fiducial cosmology. Despite $C_{l}^{\hat{\kappa}\hat{\kappa}_{\rm{null}}}$ having worse constraints than regular CMB lensing measurements by construction, we discuss how $C_{l}^{\hat{\kappa}\hat{\kappa}_{\rm{null}}}$ uniquely probes high-redshift physics. We show how such a probe could serve as a model-independent test of non-standard time evolution, illustrating such a test by constructing an {\it ad hoc} cosmology with deviations on the scale of the current Hubble tension and the $\sigma_{8}$ tension \cite{H0_tension_review,Poulin_2023,douspis2019tension}.  

\subsection{\label{sec:fisher} General Fisher Formalism}

The Fisher information matrix captures how much information an observable, $\mathcal{O}$, measured to some precision, carries about a set of model parameters that are grouped into a vector $\boldsymbol{\theta}$. The elements of the Fisher matrix are given by
\begin{equation}\label{eq:Fisher Information}
    F_{ij} = \sum_{l} \frac{1}{\sigma^2_l} \frac{\partial \mathcal{O}_l}{\partial\theta_{i}}\frac{\partial \mathcal{O}_l}{\partial\theta_{j}}
\end{equation}
where $i$ and $j$ index the parameters in the model, $l$ indexes each measured mode of the observable, and $\sigma_l$ is the error on the measurement of that mode. The Fisher matrix is the inverse of the covariance matrix, $\mathbf{F}^{-1} = \mathbf{C}$, for the set of model parameters. In our case, we perform two independent forecasts, one with $\mathcal{O} = \{C_{l}^{\hat{\kappa}\hat{\kappa}_{\rm{null}}}\}$ (the forecast of interest) and one with $\mathcal{O} = \{ C_{l}^{\hat{\kappa}\hat{\kappa}}\}$ (which serves as a reference). We compute these spectra using Eqs. \eqref{eq:CMB_spectrum} and \eqref{eq:CMBxNull_spectrum} and take their numerical derivatives using finite differences. The Fisher matrix encodes Gaussian parameter uncertainties; by the Cramer-Rao bound, this provides an optimistic approximation of the true posterior distributions. In reality, the true uncertainties may be larger than those modeled due to other systematics and may be non-Gaussian.

In addition to the parameter covariance matrix, one can compute the bias on each parameter in the presence of a systematic. Given some observable $C_{l}^{\rm{obs}}$ that contains the the signal of interest, $C_{l}$, and also some systematic contaminant, $C_{l}^{\rm{cont}}$, the total observed quantity is given by
\be
 C_{l}^{\rm{obs}}= C_{l} +C_{l}^{\rm{cont}}.
\ee
Following Ref. \cite{Huterer_Takada_2005} the $i$th component of the parameter bias vector $\mathbf{b}$ is given by 
\be\label{eq:param_bias}
b_{i} = \langle \hat{\theta}_{i} \rangle - \langle \theta^{\rm{true}}_{i} \rangle = \sum_{j}(\mathbf{F}^{-1})_{ij} \mathbf{B}_{j}
\ee
where ${\boldsymbol{\hat{\theta}}}$ contains the best fit parameter values, $\boldsymbol{\theta^{\rm{true}}}$ is vector containing the true underlying values, and $\mathbf{B}_j$ is
\be\label{eq:bias_vector}
\mathbf{B}_{j} = \sum_{l} \frac{C^{\rm{cont}}_{l}}{\sigma_{l}^{2}} \frac{\partial C_l}{\partial\theta_{j}}.
\ee

In the following section, we present the results of a Fisher forecast for the Next Generation and Futuristic nulling scenarios and compare it to constraints from regular CMB lensing measurements. Following this, in Sections~\ref{sec:H0} and \ref{sec:As} we compute the parameter bias vector in a slightly unusual application of the formalism: we consider the case where an incorrect cosmological model results in a ``theory systematic" that perturbs the inferred parameter values.

\subsection{\label{sec:Fisher_results_base Case} Concordance Cosmology}

Here we present the results of the Fisher forecast with respect to a concordance model of cosmology whose parameters and fiducial values are summarized in Table \ref{tab:params}.  In addition to the independent model parameters used to define the fiducial cosmology, we also compute four derived quantities: $\Omega_{\rm{m}}$, $\sigma_8$, $S_{8} \equiv \sigma_8 (\Omega_{\rm{m}}/0.3)^{0.5}$, and $S^{\rm{CMBL}}_{8} \equiv \sigma_8 (\Omega_{\rm{m}}/0.3)^{0.25}$.  The quantity $\sigma_8$ is the root-mean-squared variance of density perturbations on 8 $h^{-1}$ Mpc scales, and is given by
\be\label{eq:sigma8}
\sigma_{8}^{2} = \int_0^\infty \frac{k^{2}dk}{2\pi^{2}} P_{m}^{\rm lin}(k) \left[\frac{3 j_{1}(kR)}{kR}\right]^{2}
\ee
where $P_{m}^{\rm lin}(k)$ is the matter power spectrum at $z=0$ assuming {\it linear} theory, $R \equiv 8 \, h^{-1}$ Mpc, and $j_1$ is the first-order spherical Bessel function of the first kind. 

\begin{figure*}[ht]
\includegraphics[width=\textwidth]{ 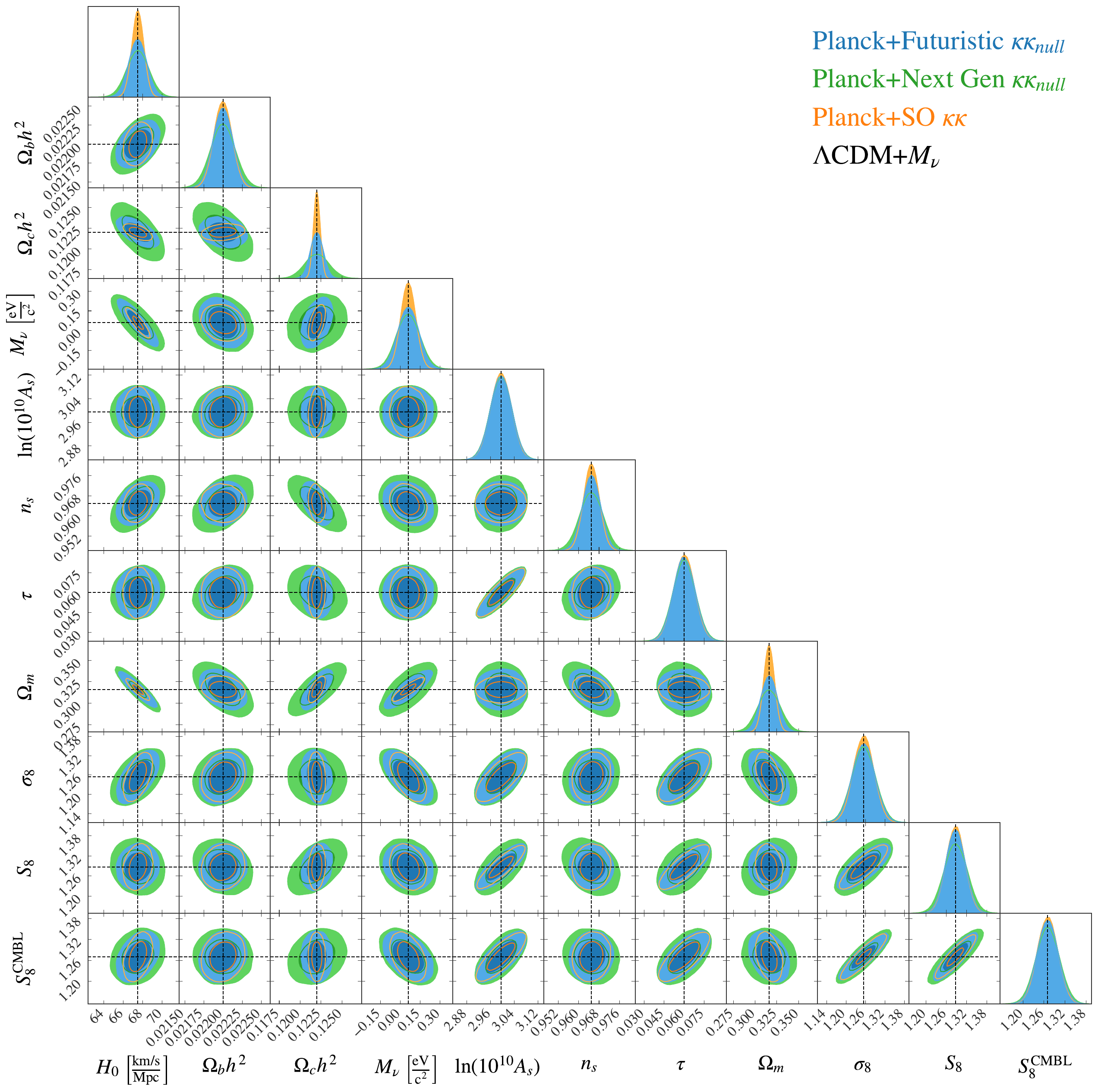}
\caption{Forecasted posterior distributions for the concordance cosmology scenario of Section~\ref{sec:Fisher_results_base Case} from SO lensing (orange), LIM-nulling in the Futuristic scenario (blue), and LIM-nulling in the Next Generation scenario (green). The dark inner region of the contours indicate the 68\% confidence region while the light coloured outer contours denote the 95\% confidence region. The dashes black lines denote the fiducial parameter values. All contours share the same Planck prior on $\Lambda$CDM + $M_{\nu}$.}
\label{fig:corner_fid}
\end{figure*}

In Fig. \ref{fig:corner_fid}, the posterior distributions are shown. In orange, the constraints from regular CMB lensing measurements forecast for SO are plotted. The green contours show the constraints from the Next Generation nulling scenario and while the Futuristic nulling scenario is in blue. \textit{Planck} 2015 priors excluding \textit{Planck} lensing have been applied to all cases \cite{Planck}. The black dashed lines mark the fiducial values of the model.
\begin{table}[b]
\centering
\caption{Model parameters and their fiducial values.}
\begin{tabular}{ccc}\toprule
Parameter         & Definition                      & Fiducial Value                          \\\midrule
$H_0$             & Hubble constant [km/s/Mpc]                & 67.5                           \\
$\Omega_{b}h^{2}$ & Fractional baryon density~      & 0.022                                   \\
$\Omega_{c}h^{2}$ & Fractional dark matter density~ & 0.120                                  \\
$M_{\nu}$         & Sum of the neutrino masses [eV/c$^2$]      & 0.06                                    \\
$A_{s}$           & Primordial fluctuation amplitude            & $2\times 10^{-9}$ \\
$n_{s}$           & Spectral index                  & 0.965                                   \\
$\tau$            & CMB Optical depth   & 0.06        \\\bottomrule                             
\end{tabular}
\label{tab:params}
\end{table}

When comparing the CMB lensing contours to the nulling ones, it is immediately obvious that the LIM-nulling probes contain less information than that of regular CMB lensing. This is entirely expected, since nulling by construction removes the low redshift information from the CMB convergence. This low-redshift information does of course have constraining power. Even with this being the case, both LIM-nulling measurements do add non-negligible information to the {\it Planck} prior and still provide comparable constraints to CMB lensing. For both CMB and CMB $\times$ LIM-nulling cases, $A_s$ and $\tau$ are not well constrained; they trace the prior. The parameters to which they are most sensitive are $\Omega_{c}h^{2}$, $H_{0}$, and $M_{\nu}$, in that order.  As for the derived quantities ($\Omega_m$, $\sigma_{8}$, $S_{8}$, and $S^{\rm{CMBL}}_{8}$), one does see some hints that select combinations of $\Omega_m$ and $\sigma_8$ are better constrained by our nulling estimator---as is usually the case for lensing. However, we caution that the Fisher formalism is not equipped to fully capture the shapes of degenerate joint posteriors between parameters. In order to make more definitive claims about these parameters, a full sampling of the posterior (e.g., via Markov Chain Monte Carlo techniques) would be more appropriate, and so we omit these parameters from the analyses in subsequent sections.

It may seem on the face of it that LIM-nulling is not worth the effort, given that it requires one to make high significance detections of not just one, but three cosmological probes---just to obtain constraints on $\Lambda$CDM parameters that are less competitive (albeit comparable) to those of CMB lensing! Yet, it is important to appreciate that CMB lensing and LIM-nulling lensing are not measuring the same thing. The power of LIM-nulling is that it is a clean probe of the high-redshift universe exclusively. To illustrate this, we forecast how the constraints of these probes differ in a universe in which we have an unexpected time evolution of parameters.

\subsection{\label{sec:H0} Early- and Late-Time Parameter Consistency: The Hubble Parameter}

Here we construct mock lensing data in a universe that obeys a cosmology which deviates from standard $\Lambda$CDM. Our toy model is inspired by the Hubble tension. When computing the lensing convergence spectrum (i.e. Eq. \ref{eq:CMB_spectrum}), for the integration steps where $ z > z_{\rm{null}}$ the value of $H_0$ that enters into $H(z) = H_0 \sqrt{\Omega_m (1+z)^3 + \Omega_{\Lambda}} $ and into the initialization of \texttt{CAMB} to obtain the matter power spectrum is  67.7 km/s/Mpc. This means that for $ z > z_{\rm{null}}$ the matter power spectrum evolves as usual in a $\Lambda$CDM cosmology and \textit{if} evolved all the way to $z = 0$ \textit{would} reach the value 67.7 km/s/Mpc. When evaluating integration steps where $ z < z_{\rm{null}}$ the same procedure is followed, but the value of $H_0$ that enters into $H(z)$ and into the initialization of \texttt{CAMB} to obtain the matter power spectrum is  72 km/s/Mpc. This model can also be described as a scenario where the growth factor, $D(z)$, and therefore the amplitude of $P(k)$ undergoes a sudden change at $z = z_{\rm{null}}$.

The two values of $H_0$ that were considered,  67.7 km/s/Mpc and 72 km/s/Mpc, constitute the discrepancy between low- and high-redshift measurements of the Hubble constant \cite{H0_tension_review}. This Hubble tension remains one of the outstanding problems of the last decade. Some argue that a time evolving cosmology may be to blame \cite{Benetti_2019,Benetti_2021}. If this were the case, LIM-nulling may be able to help elucidate this mystery as it contains no information about the late-time evolution of the matter density field, allowing a clean measurement of what happens at high redshifts. To be clear, we do not argue that the model proposed here is a genuine solution to the Hubble tension, nor that the LIM-nulling technique is destined to detect the tension. We are simply using this commonly known open cosmological problem as inspiration for how one might look for parameter consistency between early- and late-time measurements.


We fit CMB lensing measurements, which are sensitive to both values of $H_{0}$, and CMB $\times$ LIM-nulling measurements, which are only sensitive to the high redshift value of $H_{0}$, to the same fiducial model in which the Hubble parameter has not undergone an abrupt shift and therefore has one $H_{0}$ parameter. After all, one's initial null hypothesis in a real data analysis pipeline will likely not assume any abrupt shifts in cosmological parameters. Performing parameter fits using this incorrect cosmological model, we expect the CMB lensing measurements to be biased. We quantify the parameter biases using Eq. \eqref{eq:param_bias} and Eq. \eqref{eq:bias_vector}, treating the systematic contaminant to be the {\it difference} between the true lensing contribution from $0 < z < z_{\rm{null}}$ and the incorrect contribution assuming a single value of $H_0$.

\begin{figure*}[ht]
\includegraphics[width=\textwidth]{ 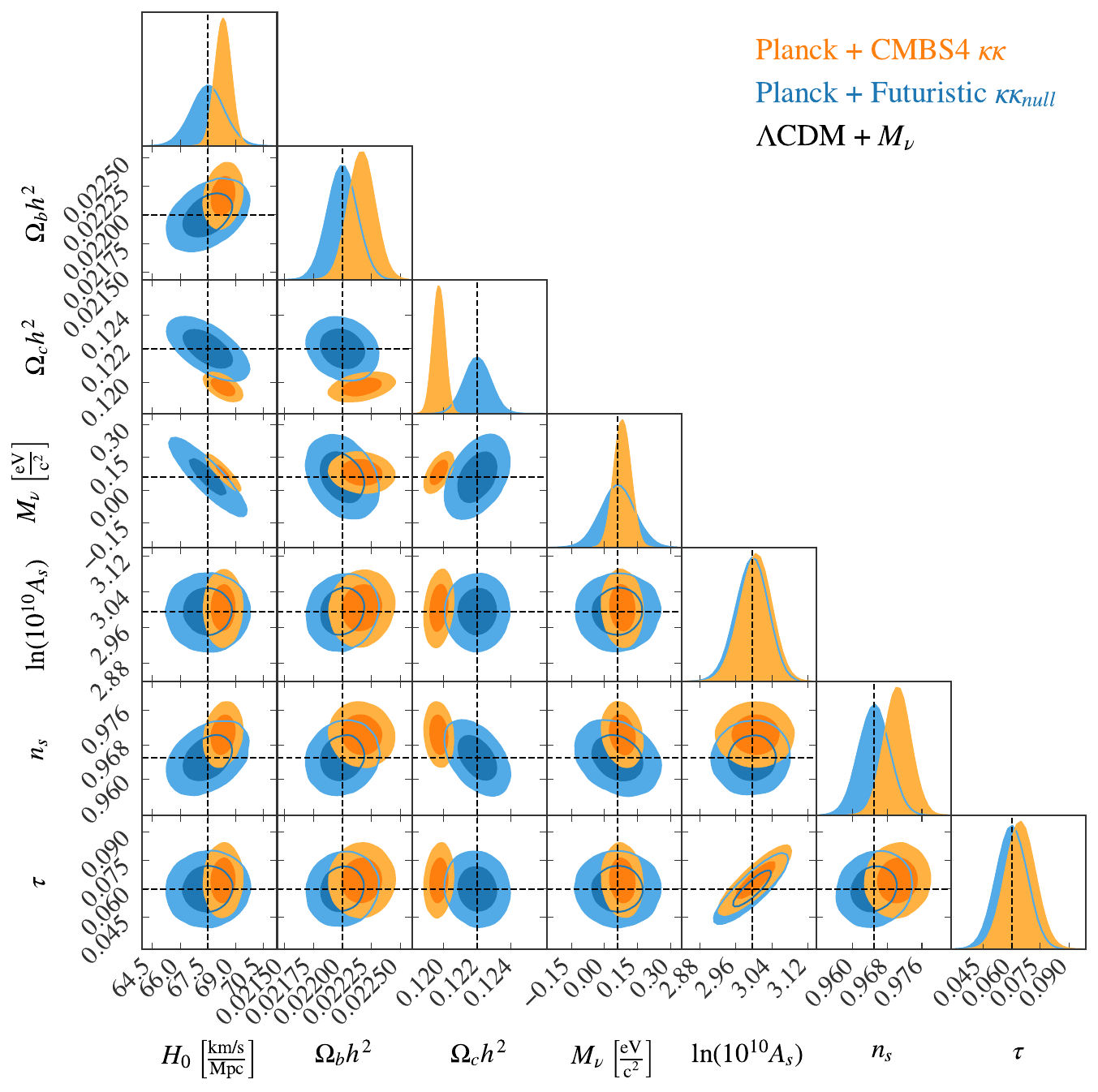}
\caption{Forecasted posterior constraints from the abruptly evolving $H(z)$ cosmology of Section~\ref{sec:H0}. Constraints are from CMBS4 lensing (orange) and LIM-nulling in the Futuristic scenario (blue). The dark inner region of the contours indicate the 68\% confidence region while the light coloured outer contours denote the 95\% confidence region. The dashes black lines denote the fiducial parameter values. These contours showcase a tension between the CMB lensing measurements and the CMB $\times$ LIM-nulling measurements. Both contours share the same Planck prior on $\Lambda$CDM + $M_{\nu}$. Since LIM nulling is sensitive only to the high-redshift universe, a comparison between the contours allows for model-independent tests of unexpected differences between high and low redshifts.}
\label{fig:corner_H0}
\end{figure*}

In Fig. \ref{fig:corner_H0}, we show the constraints on CMB lensing from CMBS4 and on LIM-nulling measurements from our Futuristic scenario. By construction, the constraints from the $C_{l}^{\hat{\kappa}\hat{\kappa}_{\rm{null}}}$ fit contain no bias with respect to the fiducial model parameters. This is not the case for the constraints from CMB lensing. It is immediately clear that there is a discrepancy between the constraints from CMB lensing and those from LIM-nulling. Such measurements would constitute a tension that provides evidence that the behaviour underlying the CMB lensing data is better described by another model.  And indeed, the CMB lensing data contains information about the universe before and after the abrupt change to $H(z)$, while the model does not. 

The parameter for which the tension is the strongest is $\Omega_{c}h^{2}$ at just over $2 \sigma$. Recall that this parameter sees the most improvement when lensing data is added to the \textit{Planck} prior. It is also worth commenting on the role the prior plays in this analysis. If one had some credence that the universe did not follow $\Lambda$CDM, one may choose to relax the \textit{Planck} prior as has been done for a number of weak lensing analyses. We found that the prior covariance can be inflated by an overall factor of 10 and still yield a $\sim 2\sigma$ tension in the inferred value of $\Omega_c h^2$ while inflating the prior by an overall factor of 20 results in a $\sim 1\sigma$ tension of $\Omega_c h^2$. When the prior is relaxed any further the tension is no longer significant. We acknowledge that in practice, more rigorous analysis would be required to claim a tension between measurements \cite{Handly_Lemos, Park_Rozo,Marshall_2006,Lemos_2021}.

An advantage of this type of measurement is that making a higher significance detection of the bias between CMB lensing and LIM-nulling can result from either improving the CMB lensing measurement \textit{or} the LIM-nulling measurement \textit{or} both. Finally, it is important to note that nowhere in the analysis did we need to make any assumption about exactly what new cosmological model \textit{did} in fact fit both data sets to detect a deviation from the standard cosmology. This constitutes a model independent test of cosmology beyond $\Lambda$CDM. 

The reader may have noticed that in order to null the low-redshift contribution to CMB lensing, one needs to compute the nulling coefficients, $\alpha$, given by Eq. \ref{eq:alpha}, which depend on the comoving distance, $\chi(z)$, and therefore the cosmology. This may seem to negate the claim of model independence if an assumption about the fiducial cosmology is needed to null in the first place. In practice, nulling can be performed through minimizing the amplitude of the LIM-nulling convergence spectrum which preserves the model independence. What nulling does is makes use of LIM lensing information by removing it from CMB lensing maps. In doing so, the CMB $\times$ LIM-nulling convergence spectrum will have a smaller amplitude than the total CMB convergence spectrum as seen in Figure \ref{fig:Cl_and_Il}. However, there is a limit to how low that spectrum can go; one can only remove as much information as is present in the LIM lensing maps. Therefore, an equivalent nulling estimator is one that minimizes the amplitude of the CMB $\times$ LIM-nulling convergence spectrum. We have explicitly verified that this minimization scheme is equivalent to the one presented in Section \ref{subsec:Nulling}. In practice, one can solve for the cosmology which yields the values of $\alpha$ that actually minimize the nulling convergence spectrum. Not only does this scheme ensure true model independence, but solving for the nulling coefficients, $\alpha$, is itself a test of cosmology and can also serve as a test for residual systematics.

\subsection{ Early- and Late-Time Parameter Consistency: Matter Fluctuation Amplitude}\label{sec:As}

Here we present a similar example to that of the previous section. We construct a scenario in which there is an abrupt change to the growth factor at $z_{\rm null}$ as a result of varying the value of $A_{s}$ for integration steps before and after $z = z_{\rm{null}}$. If one were to compute $\sigma_{8}$ with and without taking into account the changing value of $A_s$ one would obtain different values of $\sigma_8$ which corresponds roughly to the current $\sigma_8$ tension. Once again if the value of $A_s$ used to initialize the matter power spectrum for integration steps at high-$z$ is evolved to today (i.e. the standard $\Lambda$CDM scenario), one would infer $\sigma_{8} = 0.79$. If one were to take into account the abrupt evolution of the matter power spectrum at $z = z_{\rm{null}}$ (i.e. our exotic cosmology) the inferred value of $\sigma_{8} = 0.72$ \cite{Mohanty_2018_s8Tension}. Similar to the previous section, LIM-nulling would not know about the shift since it is only sensitive to the high-redshift value of $A_s$ while CMB lensing is sensitive to both values as it probes the matter density field before and after this sudden evolution. We therefore use the same parameter bias formalism to perform our forecasts.

\begin{figure*}[ht]
\includegraphics[width=\textwidth]{ 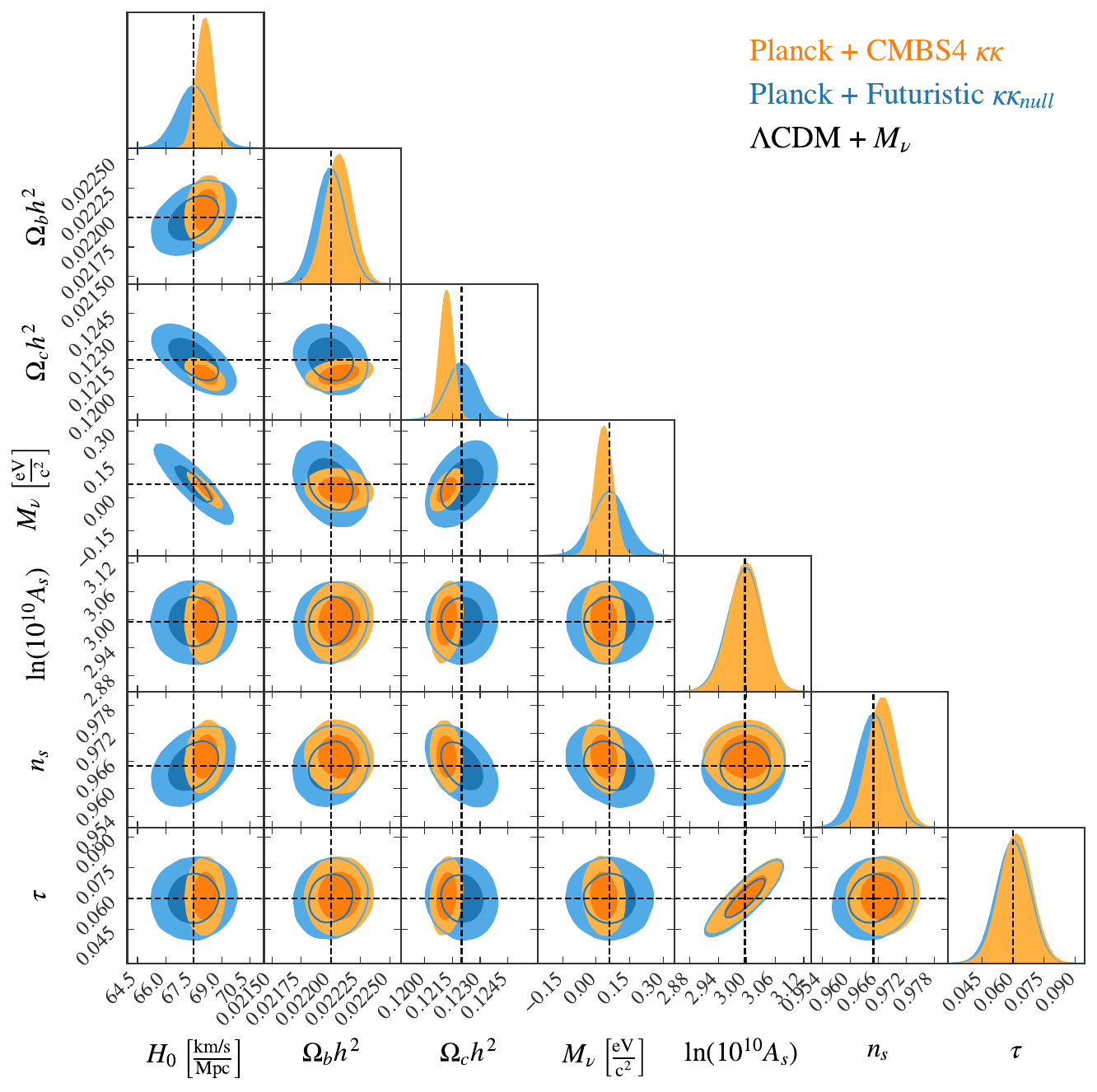}
\caption{Same as Fig.~\ref{fig:corner_H0}, except for the $A_s$-evolving cosmology of Section~\ref{sec:As}. While a slight tension is still evident, the results are less statistically significant than for the abruptly evolving $H(z)$ cosmology.}
\label{fig:corner_As}
\end{figure*}

In Fig. \ref{fig:corner_As}, we show the constraints on CMB lensing from CMBS4 and on LIM-nulling measurements from our Futuristic scenario. The constraints from $C_{l}^{\hat{\kappa}\hat{\kappa}_{\rm{null}}}$ are not biased with respect to the fiducial model parameters and, while not as extreme as the results from the previous section, there remains a slight bias to the CMB lensing constraints. There is a $\sim 1\sigma$ tension between CMB lensing and LIM-nulling measurements for $\Omega_{c}h^2$, perhaps providing a slight hint that the CMB lensing data and the LIM-nulling data are not adequately described by the same model. If one were to relax the prior by a factor of a few ( $ \lesssim 5 ) $  the slight tension is preserved but, since the tension is not as large as it was in the last section, the tension is no longer significant if relaxed any further.

\section{\label{sec:PCA} Sensitivity To the Matter Power Spectrum}

Ultimately, the CMB $\times$ LIM-nulling convergence is a kernel weighted map of the high redshift matter density field. While lensing measurements from galaxy shear and convergence and the CMB have provided the first unbiased measurements of the matter density field out to $z\sim 3$, the majority of our knowledge of the matter density field comes from the use of biased luminous tracers. LIM-nulling has the potential to reveal the unbiased high-redshift matter density field over large cosmological volumes. This leads to the additional advantage that the high-redshift matter perturbations are mode linear, making the power spectrum easier to model. We forecast the sensitivity of $C_{L}^{\hat{\kappa}\hat{\kappa}_{\rm{null}}}$ to the matter power spectrum at various length-scales and redshifts. Using the same Fisher formalism from the previous section, we define a set of parameters which are the amplitude of the matter power spectrum in various $(k,z)$ bins. In other words, we take $\boldsymbol{\theta} = [P_{m}(k_1,z_1),P_{m}(k_2,z_1),...,P_{m}(k_{\rm max},z_{\rm max}) ]$, where $k_{\rm max}$ and $z_{\rm max}$ are the maximum $k$ and $z$ values, respectively. We define 2000 $(k,z)$ bins, ranging from $0<z<1100$ and $10^{-2}$ $h$ Mpc$^{-1} < k < 1$ $h$ Mpc$^{-1}$. 

Once again, for the sake of comparison we forecast the sensitivity of both $C_{L}^{\hat{\kappa}\hat{\kappa}}$ and $C_{L}^{\hat{\kappa}\hat{\kappa}_{\rm{null}}}$ to the parameter vector $\boldsymbol{\theta}$. While CMB lensing probes length-scales across the whole redshift range, the LIM-nulling spectrum is insensitive to $z< z_{\rm{null}}$ leaving a large portion of its Fisher information matrix null. Unable to invert such a matrices, we perform principal component analysis (PCA) on the Fisher matrices. We diagonalize each of the Fisher matrices and plot the six principal eigenmodes (i.e. the eigenvectors which have the largest eigenvalues).

\begin{figure*}[h]
\includegraphics[width=0.93\textwidth]{ 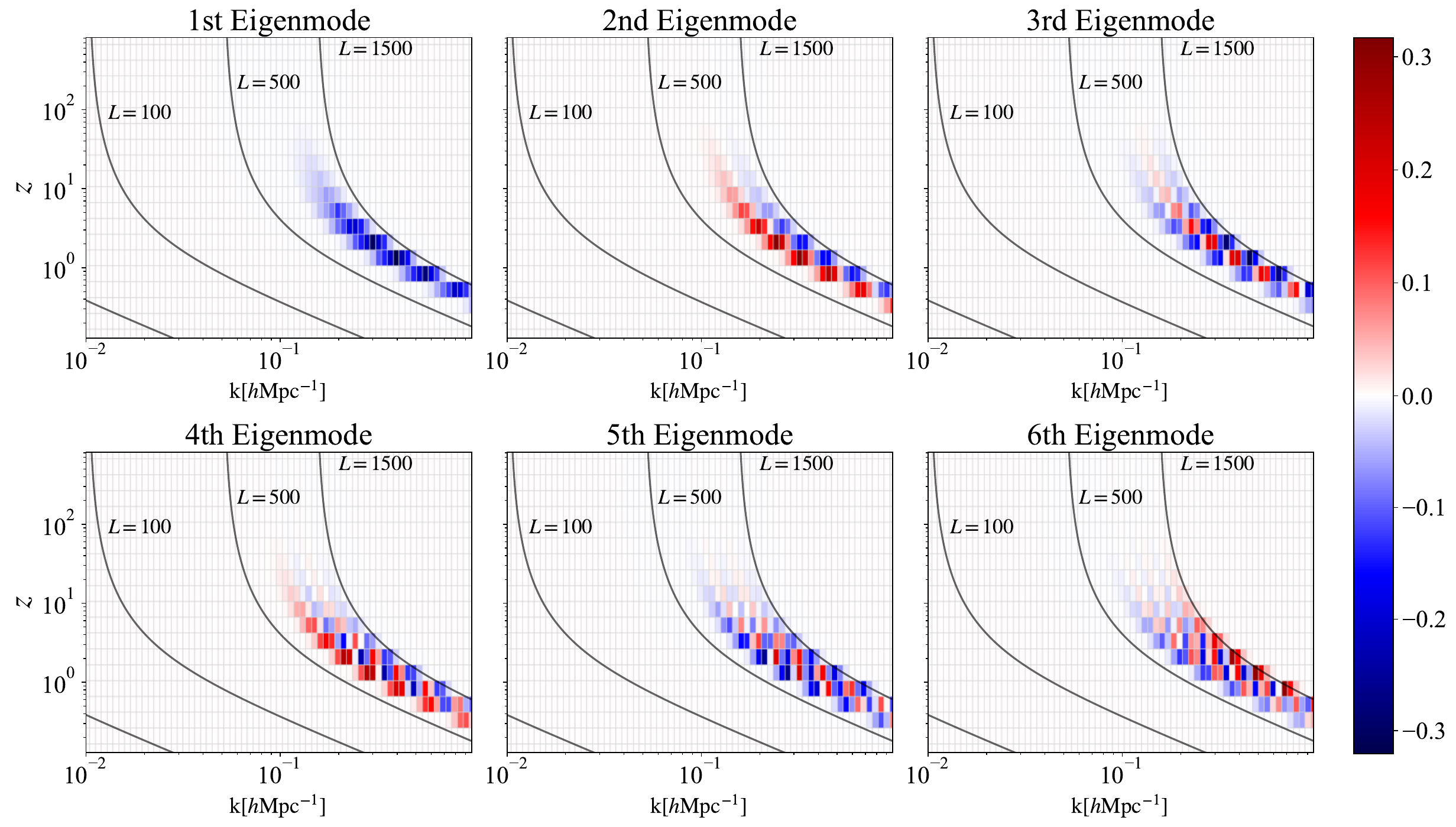}
\caption{First six principal eigenmodes of the matter power spectrum $P_m(k,z)$ for regular CMB lensing measurements from CMBS4. Lines of constant $L$ are shown in black to guide the eye, and reveal that one is essentially sensitive to an approximately fixed range of angular scales. As one goes to higher eigenmodes, one probes finer and finer features as a function of $k$. Since CMB lensing is sensitive to the integrated matter density from $z=0$ to the surface of last scattering, there is broad support as a function of redshift.}
\label{fig:PCA_S4}
\end{figure*}

\begin{figure*}[h]
\includegraphics[width=0.93\textwidth]{ 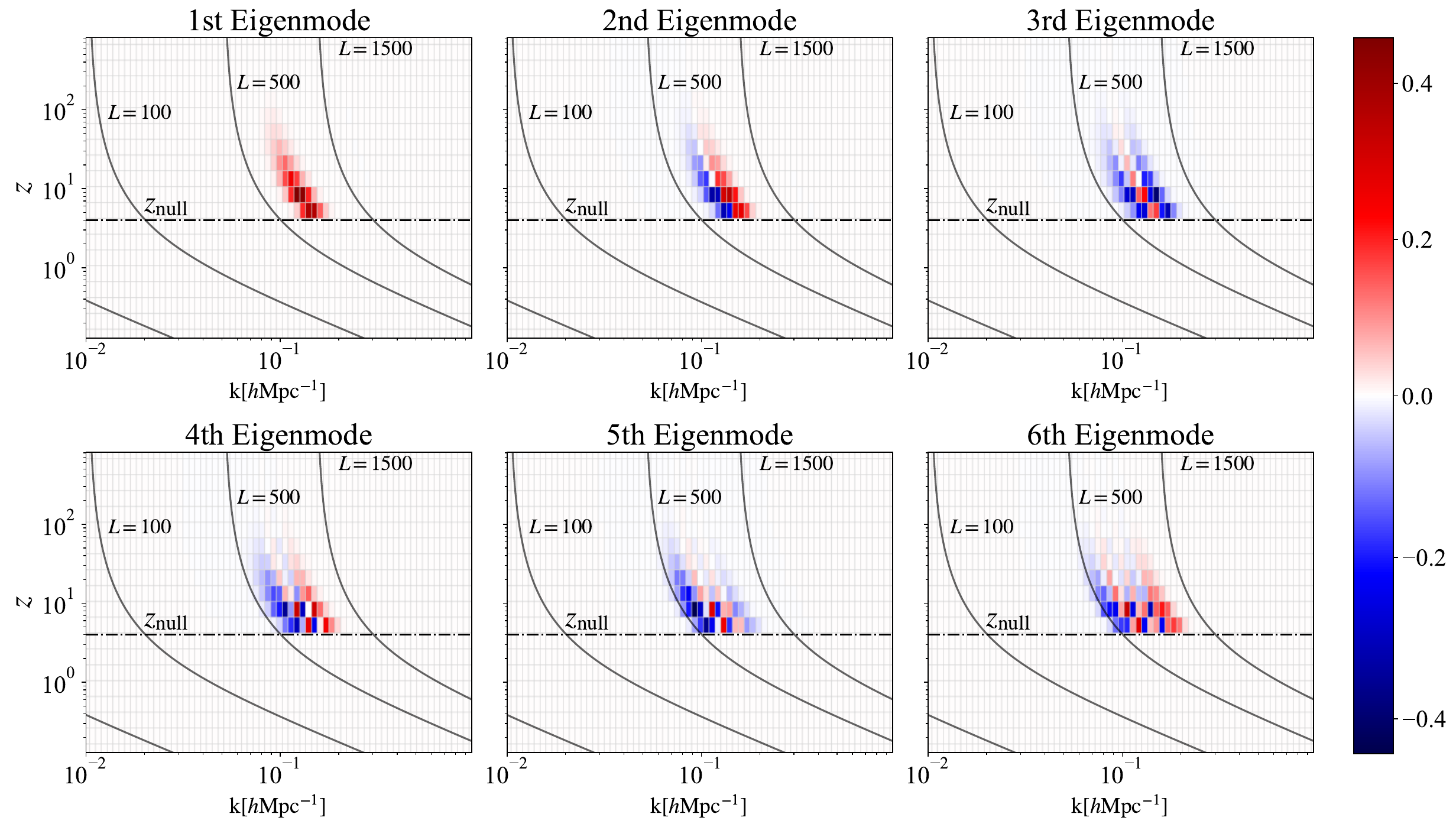}
\caption{Same as Fig.~\ref{fig:PCA_S4}, but for the CMB $\times$ LIM-nulling lensing measurements in the Futuristic scenario. By construction there is no sensitivity to redshifts below $z_{\rm null}$. The range of angular scales is also shifted slightly because of different SNR characteristics of nulling versus regular lensing measurements. The slower evolution of the mapping between $L$ and $k$ at high redshifts hints at a potential BAO measurement with nulling estimators.}
\label{fig:PCA_futureNull}
\end{figure*}

In Figs. \ref{fig:PCA_S4} and \ref{fig:PCA_futureNull}, we present the principal components for CMB lensing observed with CMBS4 and for LIM-Nulling in the Futuristic scenario, respectively. As expected, CMB lensing is sensitive to low redshift modes as its lensing kernel peaks at $z \sim 2$. The principal components are all sensitive to roughly the same range in $L$, as one can see from the lines of constant $L$ superimposed on the figures. Roughly speaking, each successive principal component probes finer and finer oscillatory modes as a function of $L$. For this CMB lensing case, the first $\sim$500 eigenmodes can be detected with with SNR $>1$.

Because the principal components follow contours of constant $L$, they contain information about many modes of wavenumber $k$ from various redshifts. This mixing of matter power spectrum modes into a single $L$ is the reason why the CMB convergence spectrum is smooth despite it being made up of the matter power spectrum which does contain acoustic peaks. This mode-mixing effect can be seen in the bottom panel of Fig. \ref{fig:Cl_and_Il}.

In the LIM-nulling case, things are appreciably different. As expected, the LIM-nulling eigenmodes peak at $z\sim z_{\rm{null}}$ since this is precisely where the nulling kernel peaks. The principal components therefore contain information about the high redshift modes of the matter power spectrum. Like in the CMB lensing case, the LIM-nulling eigenmodes trace lines of constant $L$; however: as redshift increases the relationship between $k$ and $L$ tends to one-to-one. This is driven by the fact that the relationship between angular scales and transverse comoving distances evolves slowly at high-$z$. Therefore, measuring a single $L$ in the LIM-nulling spectrum contains information about a much narrower range of power spectrum modes, which allow for one to more cleanly trace matter fluctuations.  Again, this is why BAO peaks emerge in the LIM-nulling convergence spectrum as seen in Fig. \ref{fig:Cl_and_Il}. It should also be noted the LIM-nulling eigenmodes are not simply the CMB lensing ones but with the low-$z$ portion removed. Due to the differences in the shapes of their SNR curves, these spectra probe slightly different $L$.  However, in a conceptually similar situation to the CMB lensing case, in LIM-nulling each successive eigenmode roughly corresponds to probing finer and finer features in the matter power spectrum. For the nulling case, the first $\sim$60 eigenmodes can be detected with with SNR $>1$ although with lower cumulative SNR than with CMB lensing alone, but again, the value of the nulling estimator is its clean sensitivity to the high-$z$ matter power spectrum. It is encouraging that one is able to attain a high-significance detection of principal modes of the unbiased matter power spectrum at $z \gtrsim 5$. 

The choice to constrain power spectrum modes in the range $10^{-2}$ $h$ Mpc$^{-1} < k < 1$ $h$ Mpc$^{-1}$ was no accident. Indeed this is the range of $k$ where the BAO live and where the power spectrum peaks. While we do not discuss the feasibility of measuring BAO at high redshift with LIM-nulling here, we direct the reader to Ref. \cite{LIM_nulling_BAO}.  In that work we find that it is indeed possible to constrain the BAO scale with a level of confidence comparable to other high-$z$ probes. In addition, tomographic LIM-nulling can be used as a tool to constrain other features in the matter power spectrum (not just BAO), to place upper limits on the amplitude of the matter power spectrum as a function of redshift, and to constrain the matter transfer function.

\section{\label{sec:conclusion}Conclusion}

In this paper we have shown how the CMB $\times$ LIM-nulling convergence cross-spectrum (proposed in Ref.~\cite{Maniyar_nulling}) can be used to constrain cosmology. This probe combines convergence maps of the CMB and of LIMs to exactly null out the low-redshift contribution to the CMB lensing convergence, leaving behind a direct and unbiased probe of the matter density field at high-redshift. This probe may serve to complement other high-redshift probes such as LIMs themselves, or high-redshift galaxy surveys, both of which are biased tracers of the matter density field. 

Building upon the work in Ref.~\cite{Maniyar_nulling}, we have computed the variance of the CMB $\times$ LIM-nulling convergence cross-spectrum. This enabled the SNR parameter-space studies of Sections~\ref{sec:Choice_of_Lines} and \ref{sec:Sensitivity}, which provided useful intuition for the expected sensitivity of nulling estimator measurements. Optimizing a set of experiments with some rough rules of thumb, we found that next-generation experiments may be able to make a detection.

Moving onto a set of futuristic and aspirational---but still potentially realizable---experiments, we explored the potential of nulling estimators to place constrains on standard $\Lambda$CDM parameters plus $M_\nu$. These experiments can place constraints that are comparable to those from regular CMB lensing. However, parameters derived from nulling estimators will always be slightly worse by construction, since the low-redshift information captured by regular CMB lensing does has sensitivity to cosmological parameters.

The true benefit of nulling estimators, then, is not in raw statistical sensitivity. Instead, it is in one's ability to probe high redshifts cleanly. We showed that nulling estimators can be compared to traditional CMB lensing constraints, and that such comparisons can serve as model-independent tests of cosmology beyond $\Lambda$CDM. While we do not claim that these tests can solve outstanding problems in cosmology such as the Hubble tension and the $\sigma_{8}$ tension, we use these examples to illustrate how nulling estimators can probe cosmology at early times. 

Additionally, we explicitly show that the CMB $\times$ LIM-nulling convergence probes high-redshift modes of the matter power spectrum, which can in turn be used to place limits on the matter power amplitude, the matter transfer function, and measure important features of the matter power spectrum such as BAOs. In our companion paper, Ref. \cite{LIM_nulling_BAO}, we forecast a BAO measurement with the CMB $\times$ LIM-nulling convergence and find it to be encouraging. Moreover, in all of our forecasts, we assume that one uses just one pair of LIM frequency channels to perform nulling. In future work, one can imagine taking advantage of the large bandwidth of LIM experiments, leveraging all frequency channels to perform LIM-nulling tomography. This would enable the direct study of the growth of structure.

While the prospects of using LIM lensing are promising, they do not come without serious challenges. Due to the complex astrophysical processes associated with the line emission and absorption mechanisms, LIMs are highly non-Gaussian and therefore using existing quadratic lensing estimators designed for the Gaussian CMB are sub-optimal \cite{Vafaei_2010}. It has been shown that attempting to use such estimators results in biases in the LIM lensing convergence \cite{Foreman_2018,Schann_2018}. These biases can be mitigated through various means, for instance, using filters that Gaussianize the field, or through bias-hardening, a method which makes use of our knowledge about the non-Gaussianity of LIMs in the estimation \cite{Vafaei_2010,Foreman_2018}. 

In addition, LIMs suffer from foreground contamination beyond the line interloper foregrounds included in this paper. For example, continuum foregrounds (such as the cosmic infrared background or galactic synchrotron emission) are expected to be present (although at a level that is generally considered to be less of a concern than line interlopers). One possible strategy for removing continuum foregrounds is to use their spectrally smooth nature to be removed.  For example, Ref. \cite{Yue_2015} show that the spectrally smooth far infrared (FIR) continuum foregrounds of [CII] can be removed with a negligible residual via spectral decomposition \cite{Yue_2015}. However, a fuller treatment ought to explicitly model these residuals in the context of nulling estimators.

In summary, the early epochs of the universe remain a treasure trove of cosmological information. LIM-nulling, in principle, has the potential to provide us with a clean window into the epoch of reioniziation, cosmic dawn, and even into the cosmic dark ages, allowing for unbiased measurements of the matter density field before the time of galaxy formation. While this endeavor presents a serious challenge, we have shown that the result offers a unique way to constrain cosmology beyond what is offered by existing probes, providing yet another pathway to the ultimate goal of understanding our Universe on all scales and at all redshifts.

\begin{acknowledgments}

The authors would like to thank Eiichiro Komatsu, Simon Foreman, Adélie Gorce and Giulio Fabbian for helpful discussion and comments. HF is supported by the Fonds de recherche du Québec Nature et Technologies (FRQNT) Doctoral Research Scholarship award number 315907 and acknowledges support from the Mitacs Globalink Research Award for this work. AL and HF acknowledge support from the Trottier Space Institute, the New Frontiers in Research Fund Exploration grant program, the Canadian Institute for Advanced Research (CIFAR) Azrieli Global Scholars program, a Natural Sciences and Engineering Research Council of Canada (NSERC) Discovery Grant and a Discovery Launch Supplement, the Sloan Research Fellowship, and the William Dawson Scholarship at McGill. ARP was supported by NASA under award numbers 80NSSC18K1014, NNH17ZDA001N, and 80NSSC22K0666, and by the NSF under award number 2108411. ARP was also supported by the Simons Foundation.
This research was enabled in part by support provided by Compute Canada (\href{www.computecanada.ca}{www.computecanada.ca}).
\end{acknowledgments}

\appendix

\begin{widetext}
\begin{center}
\newpage

\section{CMB \texorpdfstring{$\times$}{TEXT} LIM-Nulling Variance}\label{app:nulling_var}
\end{center}

In Ref. \cite{Maniyar_nulling}, it has been shown that the cross-correlation, $C^{\hat{\kappa}_{\rm{CMB}}\hat{\kappa}_{\rm{null}}}_{L}$, is an unbiased estimator. Here, we compute the variance of this estimator in the presence of uncorrelated Gaussian noise at the convergence map level. We also assume, as in Ref. \cite{Maniyar_nulling},
that the estimated convergence maps, $\hat{\kappa}_{\rm{CMB}}$ and $\hat{\kappa}_{\rm{null}}$, are Gaussian. 


We begin with the standard formula for the variance, i.e.,
\begin{equation}\label{eq:variance}
    \rm{var}(C^{\hat{\kappa}_{\rm{CMB}}\hat{\kappa}_{\rm{null}}}_{L}) = \langle (\hat{\kappa}_{\rm{CMB}} \hat{\kappa}^*_{\rm{null}} )^2\rangle - \langle \hat{\kappa}_{\rm{CMB}} \hat{\kappa}^*_{\rm{null}} \rangle^2.
\end{equation}
The first term can be written as 
\begin{align}\label{eq:first term}
    \langle (\hat{\kappa}_{\rm{CMB}} \hat{\kappa}^*_{\rm{null}} )^2\rangle &= \langle \hat{\kappa}_{\rm{CMB}} \hat{\kappa}^*_{\rm{null}} \hat{\kappa}_{\rm{CMB}} \hat{\kappa}^*_{\rm{null}}\rangle \nonumber \\ 
    &= \langle \hat{\kappa}_{\rm{CMB}} (\hat{\kappa}^*_{\rm{CMB}} + \alpha\hat{\kappa}^*_{2}-(1+\alpha)\hat{\kappa}^*_{1})\hat{\kappa}_{\rm{CMB}} (\hat{\kappa}^*_{\rm{CMB}} + \alpha\hat{\kappa}^*_{2}-(1+\alpha)\hat{\kappa}^*_{1}) \rangle \nonumber \\
    &=\langle\hat{\kappa}_{\rm{CMB}} \hat{\kappa}^*_{\rm{CMB}}\hat{\kappa}_{\rm{CMB}} \hat{\kappa}^*_{\rm{CMB}}\rangle + 2\alpha \langle\hat{\kappa}_{\rm{CMB}} \hat{\kappa}^*_{\rm{CMB}}\hat{\kappa}_{\rm{CMB}} \hat{\kappa}^*_{\rm{2}}\rangle -2(1+\alpha)\langle\hat{\kappa}_{\rm{CMB}} \hat{\kappa}^*_{\rm{CMB}}\hat{\kappa}_{\rm{CMB}} \hat{\kappa}^*_{\rm{1}}\rangle \nonumber \\ 
    &+\alpha^2\langle\hat{\kappa}_{\rm{CMB}} \hat{\kappa}^*_{\rm{2}}\hat{\kappa}_{\rm{CMB}} \hat{\kappa}^*_{\rm{2}}\rangle  -2\alpha(1+\alpha) \langle\hat{\kappa}_{\rm{CMB}} \hat{\kappa}^*_{\rm{2}}\hat{\kappa}_{\rm{CMB}} \hat{\kappa}^*_{\rm{1}}\rangle +(1+\alpha)^2\langle\hat{\kappa}_{\rm{CMB}} \hat{\kappa}^*_{\rm{1}}\hat{\kappa}_{\rm{CMB}} \hat{\kappa}^*_{\rm{1}}\rangle,
\end{align}
where here $\hat{\kappa}_{1}$ and $\hat{\kappa}_{2}$ are the Fourier transform of the estimated convergence maps from LIMs at comoving distances $\chi_1$ and $\chi_2$ where $\chi_1 < \chi_2$ and the asterisks denote the complex conjugate. Each term in this expression can be evaluated using the fourth-order moment relation
\begin{equation}\label{eq:fourth-order moment}
    \langle x_1 x_2 x_3 x_4 \rangle = \langle x_1 x_3 \rangle\langle x_2 x_4 \rangle + \langle x_1 x_2 \rangle\langle x_3 x_4 \rangle + \langle x_1 x_4 \rangle\langle x_2 x_3 \rangle
\end{equation}
where $x_1,x_2,x_3,x_4$ are Gaussian random variables with mean zero. Keeping in mind that the $i$th convergence map contains a cosmological signal, $s_i$, and is contaminated with uncorrelated Gaussian random noise with mean zero, $n_i$, we have $\kappa_i = s_i+ n_i$. Eq. \eqref{eq:first term} then simplifies to
\begin{eqnarray}\label{eq:first_term_foiled}
    \langle (\hat{\kappa}_{\rm{CMB}} \hat{\kappa}^*_{\rm{null}} )^2\rangle
    &=& 3[C^{\hat{\kappa}_{\rm{CMB}}}_{L} + N^{\hat{\kappa}_{\rm{CMB}}}_{L}]^2 + 3[2\alpha(C^{\hat{\kappa}_{\rm{CMB}}}_{L} + N^{\hat{\kappa}_{\rm{CMB}}}_{L})C^{\hat{\kappa}_{\rm{CMB}}\hat{\kappa}_{\rm{2}}}_{L}] - 3[2(1+\alpha)(C^{\hat{\kappa}_{\rm{CMB}}}_{L} + N^{\hat{\kappa}_{\rm{CMB}}}_{L})C^{\hat{\kappa}_{\rm{CMB}}\hat{\kappa}_{\rm{1}}}_{L}] \nonumber  \\
    &+&\alpha^2[2 (C^{\hat{\kappa}_{\rm{CMB}}\hat{\kappa}_{\rm{2}}}_{L})^2+(C^{\hat{\kappa}_{\rm{CMB}}}_{L} 
   + N^{\hat{\kappa}_{\rm{CMB}}}_{L})(C^{\hat{\kappa}_{\rm{2}}}_{L} + N^{\hat{\kappa}_{\rm{2}}}_{L})] -2\alpha(1+\alpha)[2C^{\hat{\kappa}_{\rm{CMB}}\hat{\kappa}_{\rm{1}}}C^{\hat{\kappa}_{\rm{CMB}}\hat{\kappa}_{\rm{2}}}+ (C^{\hat{\kappa}_{\rm{CMB}}}_{L} + N^{\hat{\kappa}_{\rm{CMB}}}_{L})C^{\hat{\kappa}_{\rm{1}}\hat{\kappa}_{\rm{2}}}] \nonumber \\
   &+&(1+\alpha)^2[2 (C^{\hat{\kappa}_{\rm{CMB}}\hat{\kappa}_{\rm{1}}}_{L})^2+(C^{\hat{\kappa}_{\rm{CMB}}}_{L} 
   + N^{\hat{\kappa}_{\rm{CMB}}}_{L})(C^{\hat{\kappa}_{\rm{1}}}_{L} + N^{\hat{\kappa}_{\rm{1}}}_{L})]
\end{eqnarray}
where $C^{\hat{\kappa}_{i}}_{L} \equiv \langle s_{i} s_{i}^* \rangle $, $C^{\hat{\kappa}_{i}\hat{\kappa}_{j}}_{L} \equiv \langle s_{i}s_{j}^* \rangle$, and $N^{\hat{\kappa}_{i}}_{L} \equiv \langle n_{i} n_{i}^* \rangle$. Similarly, using Eq. \eqref{eq:fourth-order moment} the second term of Eq. \eqref{eq:variance} is
\begin{eqnarray}\label{eq:second_term}
    \langle \hat{\kappa}_{\rm{CMB}} \hat{\kappa}^*_{\rm{null}} \rangle^2 &=& (C^{\hat{\kappa}_{\rm{CMB}}}_{L} + N^{\hat{\kappa}_{\rm{CMB}}}_{L})^2 + 2\alpha(C^{\hat{\kappa}_{\rm{CMB}}}_{L} + N^{\hat{\kappa}_{\rm{CMB}}}_{L})C^{\hat{\kappa}_{\rm{CMB}}\hat{\kappa}_{\rm{2}}}_{L} -2(1+\alpha)(C^{\hat{\kappa}_{\rm{CMB}}}_{L} + N^{\hat{\kappa}_{\rm{CMB}}}_{L})C^{\hat{\kappa}_{\rm{CMB}}\hat{\kappa}_{\rm{1}}}_{L} + \alpha^2(C^{\hat{\kappa}_{\rm{CMB}}\hat{\kappa}_{\rm{2}}}_{L})^2 \nonumber\\
    &&- 2\alpha(1+\alpha)C^{\hat{\kappa}_{\rm{CMB}}\hat{\kappa}_{\rm{1}}}C^{\hat{\kappa}_{\rm{CMB}}\hat{\kappa}_{\rm{2}}} + (1+\alpha)^2(C^{\hat{\kappa}_{\rm{CMB}}\hat{\kappa}_{\rm{1}}}_{L})^2 .
\end{eqnarray}
Plugging Eq. \eqref{eq:first_term_foiled} and Eq. \eqref{eq:second_term} back into \eqref{eq:variance}, and accounting for cosmic variance we obtain
\begin{eqnarray}
\rm{var}(C_L^{\hat{\kappa}_{\rm{CMB}}\hat{\kappa}_{\rm{null}}}) &=& \left(\frac{1}{f_{\rm{sky}}(2L + 1)} \right) 2[ (C^{\hat{\kappa}_{\rm{CMB}}}_{L} + N^{\hat{\kappa}_{\rm{CMB}}}_{L})^2 + 2\alpha(C^{\hat{\kappa}_{\rm{CMB}}}_{L} + N^{\hat{\kappa}_{\rm{CMB}}}_{L})C^{\hat{\kappa}_{\rm{CMB}}\hat{\kappa}_{\rm{2}}}_{L} - 2(1+\alpha)(C^{\hat{\kappa}_{\rm{CMB}}}_{L} + N^{\hat{\kappa}_{\rm{CMB}}}_{L})C^{\hat{\kappa}_{\rm{CMB}}\hat{\kappa}_{\rm{1}}}_{L}   ] \nonumber \\&+&\alpha^2[(C^{\hat{\kappa}_{\rm{CMB}}\hat{\kappa}_{\rm{2}}}_{L})^2+(C^{\hat{\kappa}_{\rm{CMB}}}_{L} 
   + N^{\hat{\kappa}_{\rm{CMB}}}_{L})(C^{\hat{\kappa}_{\rm{2}}}_{L} + N^{\hat{\kappa}_{\rm{2}}}_{L})] -2\alpha(1+\alpha)[C^{\hat{\kappa}_{\rm{CMB}}\hat{\kappa}_{\rm{1}}}C^{\hat{\kappa}_{\rm{CMB}}\hat{\kappa}_{\rm{2}}}+ (C^{\hat{\kappa}_{\rm{CMB}}}_{L} + N^{\hat{\kappa}_{\rm{CMB}}}_{L})C^{\hat{\kappa}_{\rm{1}}\hat{\kappa}_{\rm{2}}}] \nonumber \\
   &+&(1+\alpha)^2[ (C^{\hat{\kappa}_{\rm{CMB}}\hat{\kappa}_{\rm{1}}}_{L})^2+(C^{\hat{\kappa}_{\rm{CMB}}}_{L} 
   + N^{\hat{\kappa}_{\rm{CMB}}}_{L})(C^{\hat{\kappa}_{\rm{1}}}_{L} + N^{\hat{\kappa}_{\rm{1}}}_{L})],
\end{eqnarray}
which is precisely Eq.~\eqref{eq:nulling_var}.


\end{widetext}


\bibliographystyle{prsty}

\bibliography{apssamp}

\end{document}